\documentclass[12pt,preprint]{aastex}

\newcommand{\lsun}    {L_{\odot}}
\newcommand{\msun}    {M_{\odot}}

\newcommand{\etal}{et al.}
\newcommand{\msunyr}{M_{\odot}~{\rm yr}^{-1}}
\newcommand{\mdot}{\dot M}  

\slugcomment{Printed \today}

\shorttitle{L1551 IRS 5 Binary System}
\shortauthors{}

\received{2002 July 11}
\begin{document}

\title{A Comprehensive Study of the L1551 IRS 5 Binary System}

\author{Mayra Osorio\altaffilmark{1}, 
 Paola D'Alessio\altaffilmark{2}, James Muzerolle\altaffilmark{3}, 
Nuria Calvet\altaffilmark{1}, Lee Hartmann\altaffilmark{1}}

\altaffiltext{1}{Harvard-Smithsonian Center for Astrophysics, 60 Garden St.,
Cambridge, MA 02138, USA; mosorio@cfa.harvard.edu, 
 p.dalessio@astrosmo.unam.mx, jamesm@as.arizona.edu, 
ncalvet@cfa.harvard.edu, hartmann@cfa.harvard.edu  }
\altaffiltext{2}{Instituto de Astronom\'{\i}a,
UNAM, Ap. Postal 70-264, Cd. Universitaria, 04510 M\'exico D.F.,
M\'exico}
\altaffiltext{3}{Steward Observatory, 
University of Arizona, 933 North Cherry Avenue, 
 Room N204, Tucson, AZ 85721-0065}

\begin{abstract}

We model the Class I source L1551 IRS 5, adopting a
flattened infalling envelope surrounding a binary disk system
and a circumbinary disk. 
With our composite model, we  calculate self-consistently the spectral
energy distribution of each component of the  L1551 IRS 5 system, 
using additional constraints from
recent observations by ISO, the water ice feature from  observations
 with  SpeX, the SCUBA extended spatial 
brightness distribution at sub-mm wavelengths, and 
the VLA spatial intensity distributions at 7 mm of the binary disks.  
 We analyze the sensitivity of our results to the various parameters
involved.  Our results show that a flattened envelope collapse model 
  is required to explain  
 simultaneously the large scale fluxes  and the water ice and 
silicate features. On the other hand, we find that
 the circumstellar disks are optically thick in the millimeter range and
are inclined so that their outer parts hide the emission along the line
 of sight  from their 
 inner parts.  We also find that these disks have
   lower mass accretion rates than  the infall
rate of the envelope.

\end{abstract}

\keywords{Physical data and processes: binary, disks ---  stars:
circumstellar matter,  formation  ---ISM: individual 
(L1551 IRS 5) }

\section{INTRODUCTION}
\label{introdu}

L1551 IRS 5 in Taurus (distance 140 pc; Kenyon, Dobrzycka \& Hartmann
1994) is one of the most extensively studied young stellar objects (YSOs).  
It is the prototypical embedded YSO (Strom, Strom, \& Vrba 1976) with a
massive molecular outflow (Snell, Loren \& Plambeck et al. 1980). Adams,
Lada, \& Shu (1987) identified it as a protostellar (Class I)  source with
an estimated infall rate for its envelope of $\sim 10^{-5} \msunyr$. The
near-infrared spectrum exhibits first-overtone absorption in CO,
suggesting that IRS 5 is an FU Ori object (Carr, Harvey, \& Lester 1987),
consistent with an analysis of the optical spectrum observed in scattered
light (Mundt et al. 1985 and Stocke {\etal} 1988).  Given the wide variety
of observational constraints available, L1551 IRS 5 is clearly an ideal
system for investigating the circumstellar environment that surrounds
protostars and to understand its implications for the general process of
star formation.

The spectral energy distribution (SED) of dust thermal emission in L1551
IRS 5 and spatial intensity profile at several wavelengths have been
modeled previously by several authors (Adams, Lada, \& Shu 1987; Butner et
al. 1991; Kenyon, Calvet \& Hartmann 1993, hereafter KCH93; Men'shchikov
\& Henning 1997;  White et al. 2000). Many of the models developed so far
predict SEDs with silicate absorption features much deeper than that
revealed by recent observations with the Infrared Space Observatory (ISO)  
and adopt outer envelopes smaller than indicated by sub-mm SCUBA
observations (Chandler \& Richer 2000). Some models assume simple
power-law density distributions for the circumstellar envelope without
requiring physical self-consistency.  In addition, in most of the models,
L1551 IRS 5 is considered as a single-star system, while high-resolution
interferometric observations (Rodr\'\i guez et al. 1998) indicate that IRS
5 is a binary.

In this work, we present a physically self-consistent model to investigate
the circumstellar material in L1551 IRS 5 on different scales. We assume
flattened envelope models (Hartmann et al. 1994;  Hartmann, Calvet \& Boss
1996, HCB96) for which we solve the transfer of radiation following the
methods described by KCH93 and HCB96. We further adopt viscous accretion
disks that take into account the irradiation by the envelope such as those
described by D'Alessio, Calvet \& Hartmann (1997, DCH97).  With this
composite model, we self-consistently calculate the overall SED and
spectral features, the sub-mm spatial intensity profile, the SEDs of the
inner disks, and compare with observations to constrain the envelope and
disk parameters.

This paper is organized as follows. In \S 2 we present a description of
the L1551 IRS 5 system, with a summary of the observations used in our
modeling efforts.  In \S 3 we report spectra at 2-5 microns which
constrain disk properties and ice absorption. In \S 4 we discuss the main
assumptions of our modeling.  In \S 5 we present our results and in \S 6
we discuss them. Finally, in \S 7 we present a summary and conclusions.

\section{THE L1551 IRS 5  SYSTEM}
\label{sistema}

The complexity of the inner regions of IRS 5 has been increasingly
recognized in recent years. The possibility of the central source being a
binary system was first suggested by Bieging \& Cohen (1985)  from cm-wave
observations and Looney, Mundy, \& Welch (1997) from mm-wave data.  
Detailed interferometric observations at 7 mm clearly show dust emission
in two compact elongated structures with diameters $\sim$ 20 AU, which
presumably are the circumstellar disks of a binary system (Rodr\'\i guez
et al. 1998).  Each disk may have its own ionized jet (Rodr\'\i guez
\etal~1998; Fridlund \& Liseau 1998).

On somewhat larger scales, mm-wave observations suggest an elongated
structure with scales of $\sim$ 100-400 AU (Keene \& Masson et al. 1990;
Lay et al. 1994; Looney et al. 1997). This structure may correspond to a
circumbinary ring or disk, (Looney et al. 1997), as in the case of GG Tau
(Dutrey, Guilloteau \& Simon 1994). A structure with size in the range of
$400<R_{\rm ext} \lesssim 1000$ AU has been inferred from millimeter and
submillimeter continuum observations (Ladd et al. 1995, Hogerheijde et al.
1997), but it is difficult to know if this structure corresponds to the
circumbinary disk or to the inner part of the extended envelope.

Sub-mm images of broad dynamic range obtained with SCUBA (Chandler \&
Richer 2000)  as well as millimeter continuum observations with IRAM
(Motte \& Andr\'e 2001)  reveal dust thermal emission at large scales
extending $\sim$ 10000 AU in L1551 IRS 5. The large-scale structure of the
envelope is flattened (Saito et al. 1996; Momose et al. 1998; Fridlund et
al 2002). Observations carried out by Ladd \etal\ (1995) at submillimeter
continuum reveal a large-scale cross-shaped pattern with size $\sim$ 1400
AU which may be caused by the outflow carving out a section of the
circumstellar material, consistent with the observation of scattered light
in the outflow lobe (Strom \etal~1976; Stocke \etal\ 1988).

In summary, the L1551 IRS 5 system appears to contain two protostars, each
surrounded by a circumstellar disk, both encircled by a circumbinary disk,
and all disks surrounded by an extended infalling flattened envelope.  
Our complex model contains all these components; we need therefore to
constrain parameters using the available data over a wide range of
wavelengths, as outlined above. Table 1 lists the long wavelength data
($\lambda > 300~\mu$m) we aim to model, compiled from the literature, with
references indicated. Column 3 gives the angular resolution of the
observations, and column 5 indicates the estimated radius of the observed
structure in the corresponding reference. Fluxes at $\lambda < 300~\mu$m
are taken from the compilation of KCH93. We also use spectroscopic
observations carried out by the ISO Long and Short Wavelength
spectrometers (LWS, SWS) toward IRS 5 (White et al. 2000), as well as our
scaled data from SpeX (\S 3).

\section{NEAR-IR OBSERVATIONS}

\begin{figure}
\plotone{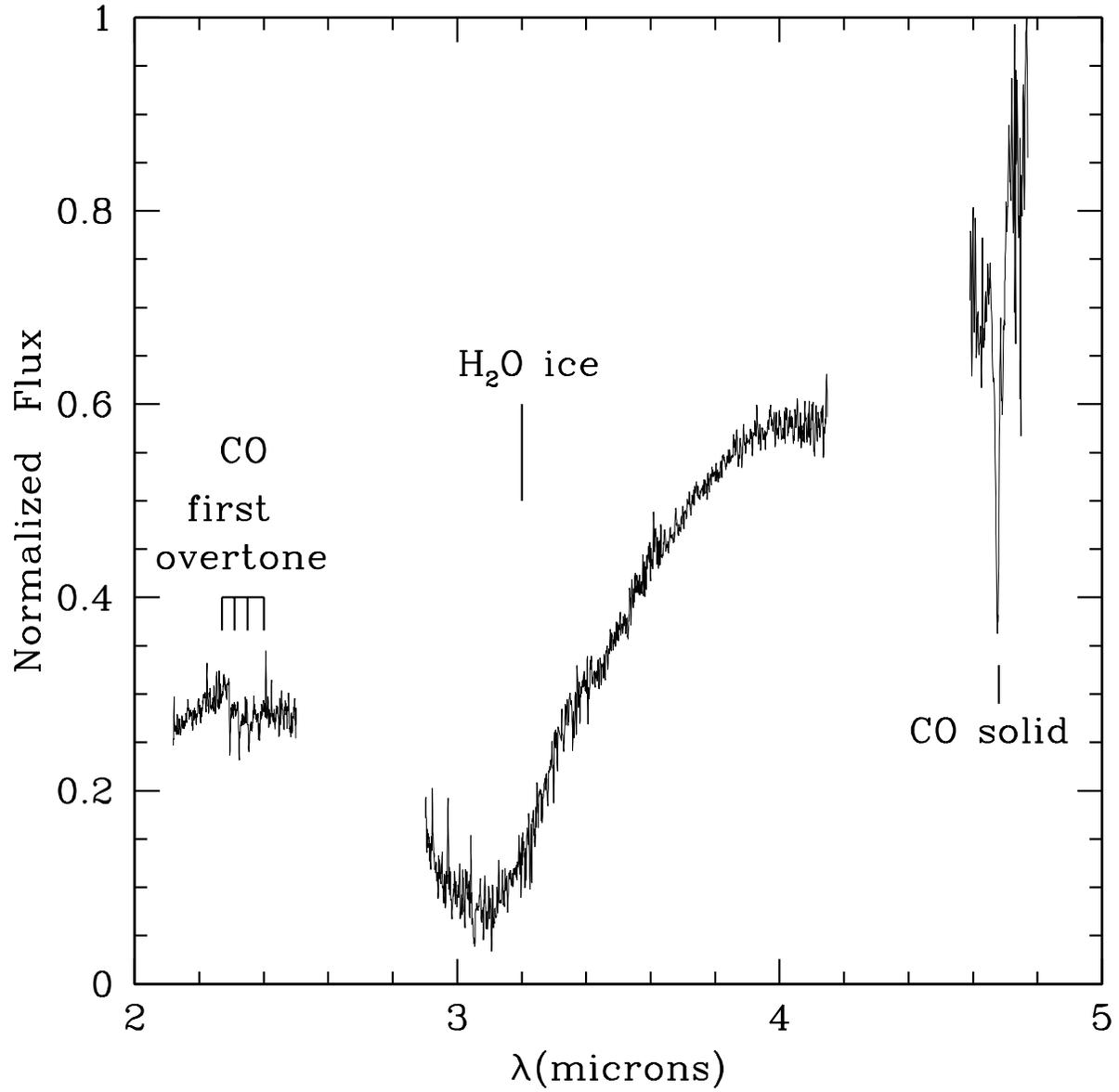}
\caption{Near-infrared spectrum of L1551 IRS 5 obtained with SpeX. 
 The telluric absorption lines have already removed. Spectral features 
mentioned in the text are marked. } 
\label{SpeX}
\end{figure}

We obtained a near-infrared spectrum of L1551 IRS 5 with SpeX (Rayner et
al. 1998) at the IRTF, on Jan. 5, 2001 (Muzerolle et al. 2002, in
preparation).  Since the spectrograph contains a cross-disperser, one can
obtain a spectrum across several bandpasses simultaneously; our spectrum
has a wavelength range of 2.1 - 4.8 microns, spanning K, L, and a portion
of M. The spectral resolution was $R\sim$~2000.  The data were taken in a
series of 10 ``ABBA" offsets, with 2 coadded 10-second exposures at each
position.  The spectrum was reduced in part using the facility IDL program
``Spextool".  Wavelength calibration was done via an arc lamp spectrum and
the sky emission lines, and is accurate to about 0.001 microns.  In order
to remove telluric absorption lines, a spectrum of a solar-type star was
taken near in time and airmass to the L1551 observation. After dividing by
the standard spectrum, the solar spectrum was used to remove residual
features due to photospheric absorption lines in the standard. The
observed spectrum is plotted in Figure \ref{SpeX}. The broad water ice
feature at $\lambda \sim 3.1 \mu$m and the first overtone bands of CO at
$\lambda \sim 2.3 \mu$m are apparent. Also noticeable is the CO solid
feature at $\lambda \sim 4.7 \mu$m.

\section{THE MODEL \label{modelo} }

\begin{figure}
\plotone{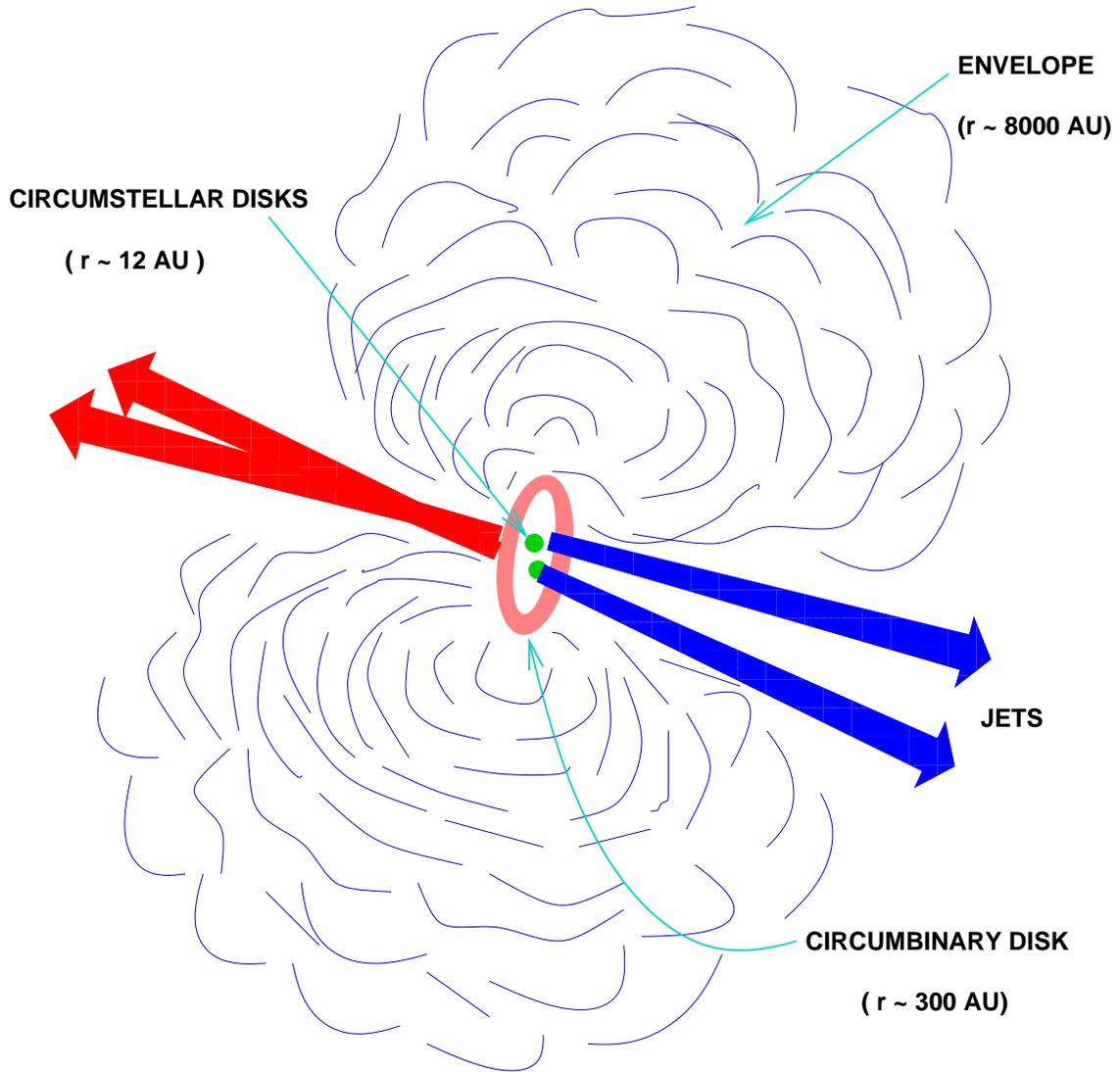}
\caption{Sketch of the geometry of the model for L1551 IRS 5} 
\label{sketch}
\end{figure}

As suggested by the observations (\S \ref{sistema}), our model includes
two circumstellar disks, a circumbinary disk, and a flattened envelope.  
A sketch of the model is show in Figure \ref{sketch}.  We describe in this
section the assumptions we use to model each component of the system.

\subsection {Envelope}
\label{modenv}

We adopt the analytic model for the flattened density distribution
resulting from the gravitational collapse of a sheet initially in
hydrostatic equilibrium developed by Hartmann et al. (1994) and HCB96,
hereafter called $\eta$ models.  These models produce envelopes with a
flat density structure at large scales, consistent with observations (\S
2). The parameters that describe the properties of the envelope models are
$\eta$, which is a measure of its asphericity ($\eta=R_{\rm out}/H$, where
$R_{\rm out}$ is the outer radius of the envelope and $H$ is the scale
height; see HCB96);  ${\rho_1}$, the density that the infalling envelope
would have at 1 AU if there were no rotation and flattening (KCH93); and
$R_c$, the centrifugal radius (Cassen \& Moosman 1981; Terebey, Shu, \&
Cassen 1984), which is the largest radius on the equatorial plane that
receives the infalling material.

The temperature structure of the envelope is calculated from the condition
of radiative equilibrium, assuming a single central source. The
spherically symmetric temperature is calculated using an angle-averaged
density, as in KCH93. The luminosity of the central source is taken as the
sum of the luminosities of the binary circumstellar disks, which are much
brighter than their central stars (\S 5.3). The spectrum of the central
source impinging on the inner radius of the envelope, taken as the dust
destruction radius, is that of a disk with luminosity equal to the total
luminosity. The temperature at the dust destruction radius is assumed to
be $\sim 1200$ K, corresponding to the sublimation temperature of
silicates in a low density environment (D'Alessio 1996). Finally, the
monochromatic emergent specific intensity is calculated by solving the
transfer equation along rays that pierce the envelope, using the
angle-dependent density distribution, as in KCH93 and HCB96. The source
function is calculated with a mean intensity which includes the direct
radiation from the central source, and the diffuse field from the
spherically symmetric calculation (cf. Calvet et al. 1994). The
monochromatic emergent flux is obtained by direct spatial integration of
the intensity.

We assume that dust in the envelope has the standard grain-size
distribution $n(a) \propto a^{-3.5}$ of the interstellar medium, with a
minimum grain radius $a_{min}$=0.005 $\mu$m and a maximum radius
$a_{max}$=0.3 $\mu$m. We use a dust mixture with olivine, organics,
troilite and water ice as it is suggested by Pollack et al. (1994, P94),
as well as a mixture with ``astronomical silicates'' and graphite
following the Draine \& Lee (1984, DL84) prescription. One of the main
differences between these mixtures is that DL84 consider that the carbon
is in graphite while P94 assume that this element is in organics.  
Furthermore, P94 include other compounds, as water and troilite. We
consider grains to be compact segregated spheres and calculate their
efficiency factor $Q_{abs}$ using a Mie scattering code (Wiscombe 1979,
see more details in D'Alessio, Calvet \& Hartmann 2001, hereafter DCH01).
Optical properties for the compounds are taken from Warren (1984),
Begemann et al. (1994) and P94.

\subsection {Circumstellar disks}

We assume that the two circumstellar disks observed at 7 mm by 
Rodr\'\i guez et al. (1998)
are $\alpha$ accretion disks irradiated by the infalling envelope. 
The structure and emission of these disks is calculated 
as described in D'Alessio (1996) and DCH97.
In short, the set of vertical structure
equations are solved for a disk with a mass accretion
rate $\mdot$ around a star with mass $M_*$ and radius $R_*$.
The heating of the disks is determined by viscous dissipation
and envelope irradiation. The viscosity $\nu$ is calculated with
the $\alpha$ prescription (Shakura \& Sunyaev 1973).
Since data exist for each disk, we
calculate the structure of each separately.

  The disks are displaced relative to the center of symmetry of the
system, thus, the envelope irradiation on each disk is not symmetric
around its own rotational axis. However, since the disks are small
compared to the size of the envelope, we approximate the backwarming to be
constant over each disk. The flux crossing the equatorial plane at a given
distance from the envelope center is calculated solving the 2D radiative
transfer throughout the envelope, as described in DCH97.  The flux
intercepted by each disk is then averaged over the disk annuli to give the
irradiation flux on the disk, $F_{irr}$, or equivalently, the irradiation
temperature, $T_{irr} = ( F_{irr}/\sigma_R)^{1/4}$, which is used as input
for the disk vertical structure calculations. The envelope model used in
these calculations is the one that gives the best fit of the SED and to
the large-scale brightness sub-mm distribution.

Since from present observations we cannot know if one of the disks is the
dominant luminosity source in the system, we assume the same mass
accretion rate and stellar parameters for both disks. We also assume that
the disks have the same dust composition than the envelope. However, to
explore the possibility of grain growth, we consider $a_{max}$ to be a
free parameter for each disk.  Other input parameters are the inclination
angle of each disk (between their rotation axis and the line of sight),
the viscosity parameter $\alpha$ (or equivalently, its mass, DCH01), and
the disk radius $R_d$.

\subsection {Circumbinary Structure}
\label{modeldiscogrande}

In order to investigate whether the flattened structure at scales of
100-400 AU resolved at millimeter and submillimeter wavelengths
corresponds to a circumbinary disk, we also included the contribution of
this component. According to DCH97, for distances larger than $\sim$ 20 AU
from its center, the temperature $T$ and surface density $\Sigma$ of a
disk surrounded by an optically thick envelope in which the heating is
dominated by envelope irradiation can be represented by power laws in
radius as $T \sim T_0 (R/100~AU)^{-0.5}$ and $\Sigma \sim \Sigma_0
(R/100~AU)^{-1}$, where $R$ is the cylindrical radius from the disk
center.  The temperature $T_0$ is fixed by the envelope irradiation at 100
AU, and we take $\Sigma_0$, the surface density at 100 AU, as a free
parameter.  The circumbinary disk has a fixed outer radius given by the
envelope centrifugal radius $R_c$ and an inner radius which is expected to
be several times larger than the maximum separation between the
circumstellar disks (Artymowicz et al. 1991).  We present models with a
grain size distribution similar to that in the extended envelope and that
in the circumstellar disks, as the two extreme possibilities for the state
of the dust in this disk.

\section{MODEL RESULTS}

In this section, we discuss the combination of parameters characterizing
each one of the components of the L1551 IRS 5 system that can reproduce
self-consistently the observational constraints. We have arranged the
section according to the interpretation of a main constraint, but it
should be kept in mind that the results for all components are
interdependent and the present
 ordering just aims to make their description easier.

\subsection{Large scale fluxes}

\subsubsection{Envelope SED}
\label{envpar}

Each range of wavelengths in the SED of the envelope is sensitive to
different parameters. The observed peak flux and its corresponding
wavelength, which occurs generally at the far infrared (IR) range, depend
mostly on the density of the envelope, which fixes the scaling $\rho_1$,
with a weak dependence upon the total luminosity of the system (see trends
of KCH93 for low mass protostars and Osorio et al. 1999, for high mass
protostars). On the other hand, the mid and near-IR fluxes are sensitive
to the $\eta$ parameter, the inclination to the line of sight $i$, and the
centrifugal radius $R_c$. The radius of the envelope is determined by the
large scale spatial intensity profiles (\S \ref{milsp}) and we take it as
$\sim$ 8000 AU.

The apparent luminosity of the system (calculated by direct integration of
the observed SED, that would correspond to the luminosity of an isotropic
source) is $L_{app} = 25 {L_{\odot}}$, and it is an indicator of the
total luminosity $L_{tot}$. For a given total luminosity and model, the
apparent luminosity depends on the inclination of the system to the line
of sight, especially in systems with a high degree of flattening. For
instance, for $\eta = 2.5$, the ratio $L_{app}/L_{tot}$ goes from $\sim$
1.4 if the system is seen pole-on, $i=0$, to $\sim$ 0.5, for edge-on
systems.  Fortunately, for the range of inclinations that produce the best
fits to the observed SEDs, $L_{app}/L_{tot} \sim 1$. Given the
approximations in our treatment, specially the assumption of spherical
symmetry for the envelope temperature distribution, we take here that
$L_{tot} = L_{app}$.

The observed flux at the peak of the SED indicates a value of
$\rho_1 \sim 4.5 \times 10^{-13}$ g cm$^{-3}$,
which also consistently results in high fluxes for the
large beam size observations in the submillimeter
and millimeter ranges (\S \ref{milsp}). 
 This value of $\rho_1$, is the maximum density that 
can produce the high millimeter fluxes
without too much extinction in the mid-infrared.
The luminosity and $\rho_1$ are similar 
to those found by KCH93 for L1551, who  considered envelopes  slightly
flattened in their inner part by  rotation as described by Terebey, 
Shu \& Cassen (1984, hereafter TSC envelopes); however those models
predict a very deep silicate feature which is not observed. 

The  recent silicate (10$\mu$m) and water ice  (3$\mu$m, 6$\mu$m) 
features observed by ISO and SpeX  place stringent limits on 
the envelope  parameters and its geometry. 
 Even in  the case of moderate inclination angles,  TSC envelopes  
predict too much extinction at short wavelengths and therefore features  
  deeper than  the observed ones. Flattened $\eta$ envelopes  with the 
same  $\rho_1$ as in TSC envelopes,  have less polar material 
and more equatorial material. So, at inclination angles to the 
 line of sight closer to the rotation axis, these  envelopes have less 
extinction than  TSC envelopes.

\begin{figure}
\plotone{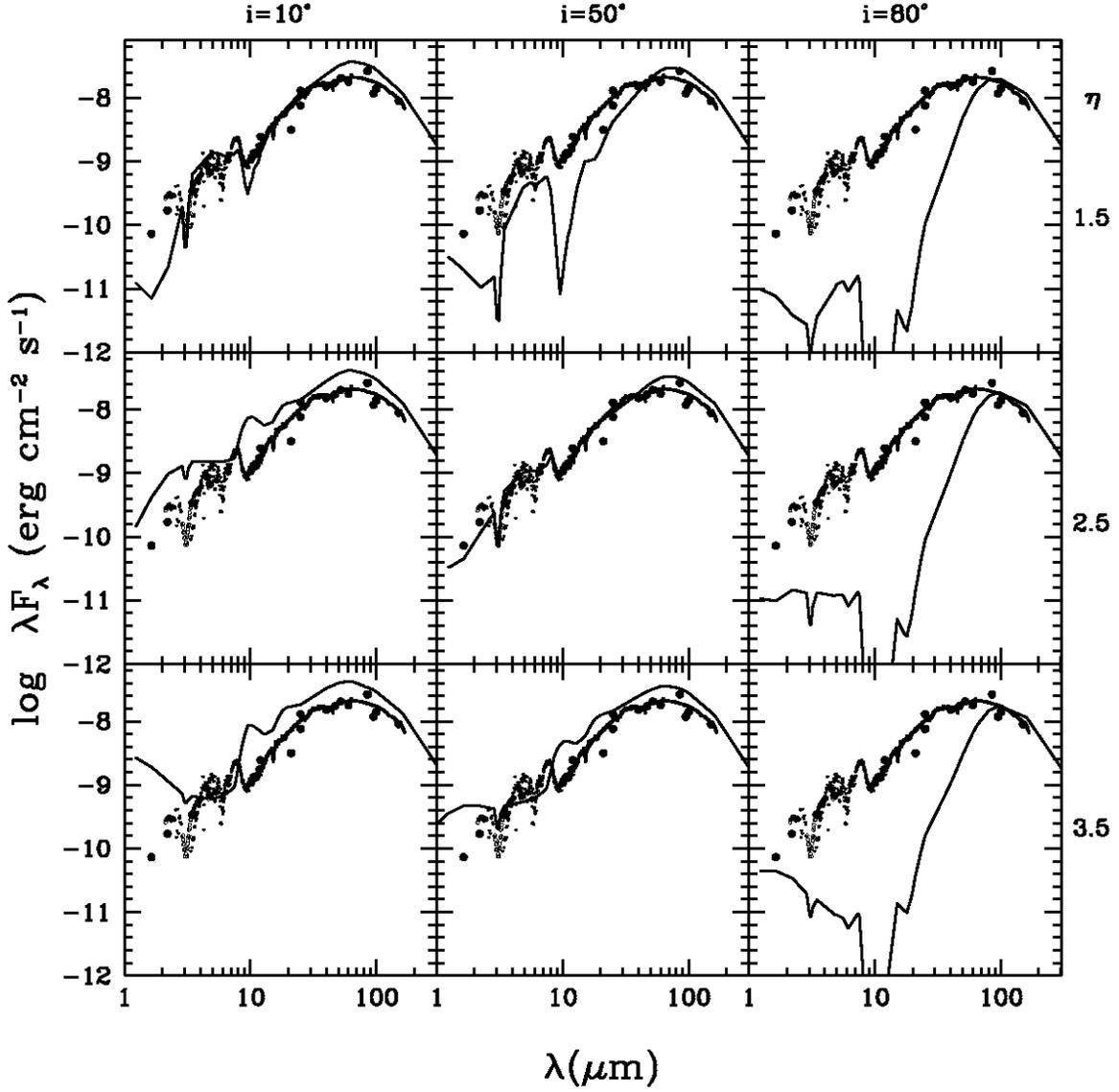}
\caption{Grid of model SEDs for inclination angles $10^{\circ}, 50^{\circ}, 
80^{\circ}$ (columns) and $\eta$= 1.5, 2.5, 3.5 (rows). All the models have
 $R_c=$300 AU, $\rho_1=4.5 \times 10^{-13}$ g cm$^{-3}$, $L=25 L_{\odot}$,
 $D=140$ pc, and an opacity with the Modified Draine \& Lee mixture 
 (see \S \ref{efectopa}). ISO (filled-small dots),  SpeX (open-small dots) 
 observations and photometry  compiled by KCH93 (filled-big dots) 
are also shown.
 } 
\label{grid1}
\end{figure}

To show the dependence of the SED on $\eta$, $i$, and $R_c$, we
present model SEDs for different values of these parameters,
with $L\*$ and $\rho_1$ fixed by the considerations above, and
compare them with observations (\S 2).
Further constraints come from other evidences.
The nearly pole-on case ($i=10^{\circ}$) is ruled out
because the outflow extends considerably in the plane of the
sky, with well-separated red- and blueshifted CO outflow lobes
(Snell \etal~ 1980). In addition, 
it is likely that the binary system formed out of lower angular momentum
material than is currently falling in; moreover, the evidence 
 for a circumbinary disk also suggests that infalling material 
 has a larger specific angular momentum than that of the binary separation.  
 For these reasons, we will consider centrifugal radii larger than the maximum
separation of the binary sources ($\sim 40$ AU).
Figure \ref{grid1} (solid line) shows the model SEDs for $R_c =$300 AU
for different inclinations and values of $\eta$.
Model SEDs in Figure \ref{grid1} were calculated
with abundances of silicates and graphite consistent with DL84 and
an abundance for water ice lower than P94 by a factor of two.
We will consider the effects of dust
opacities in more detail in \S \ref{efectopa}.

\begin{figure}
\plotone{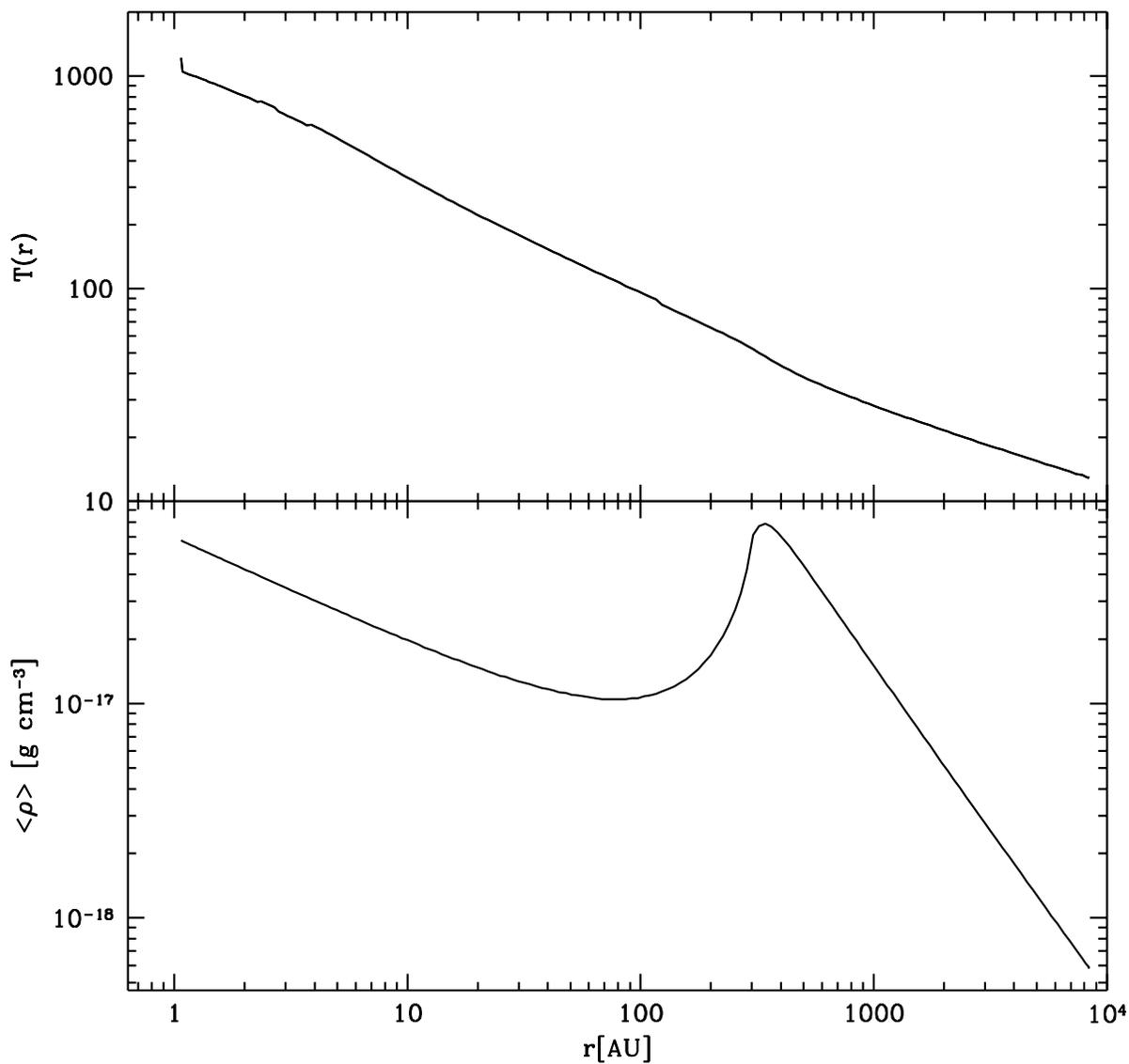}
\caption{Temperature  (upper panel) and angle-averaged density distribution
 (lower panel) of the envelope fiducial model: $i \sim 50^{\circ}$, 
$\eta \sim 2.5$ and $R_c$=300 AU (see \S \ref{envpar}). Inside the centrifugal 
 radius the angular momentum of the infalling material causes departures from 
 the radial free fall.} 
\label{tempdensi}
\end{figure}

In general, all  models for high inclination angles ($\sim$ 80$^{\circ}$)
predict too little near and mid-infrared emission, 
regardless of the degree of flattening ($\eta$ parameter).
As discussed above, models with roughly  spherical envelopes 
 ($\eta=1.5$, similar to the 
case of TSC envelopes)  have too much extinction 
along of the line of sight 
and therefore exhibit   deeper  absorption features than 
models which assume flatter envelopes. 
Thus, models for flattened envelopes and with  
moderate inclinations explain better the observed SED; in particular 
 the best fits are obtained  at inclination angles  
 $i \sim 50^{\circ}$ and $\eta \sim 2.5$ for
 the case of $R_c$=300 AU, which we adopt as our fiducial model. 
 The angle-averaged density and temperature distributions of this model 
 are shown in Figure \ref{tempdensi}.

\begin{figure}
\plotone{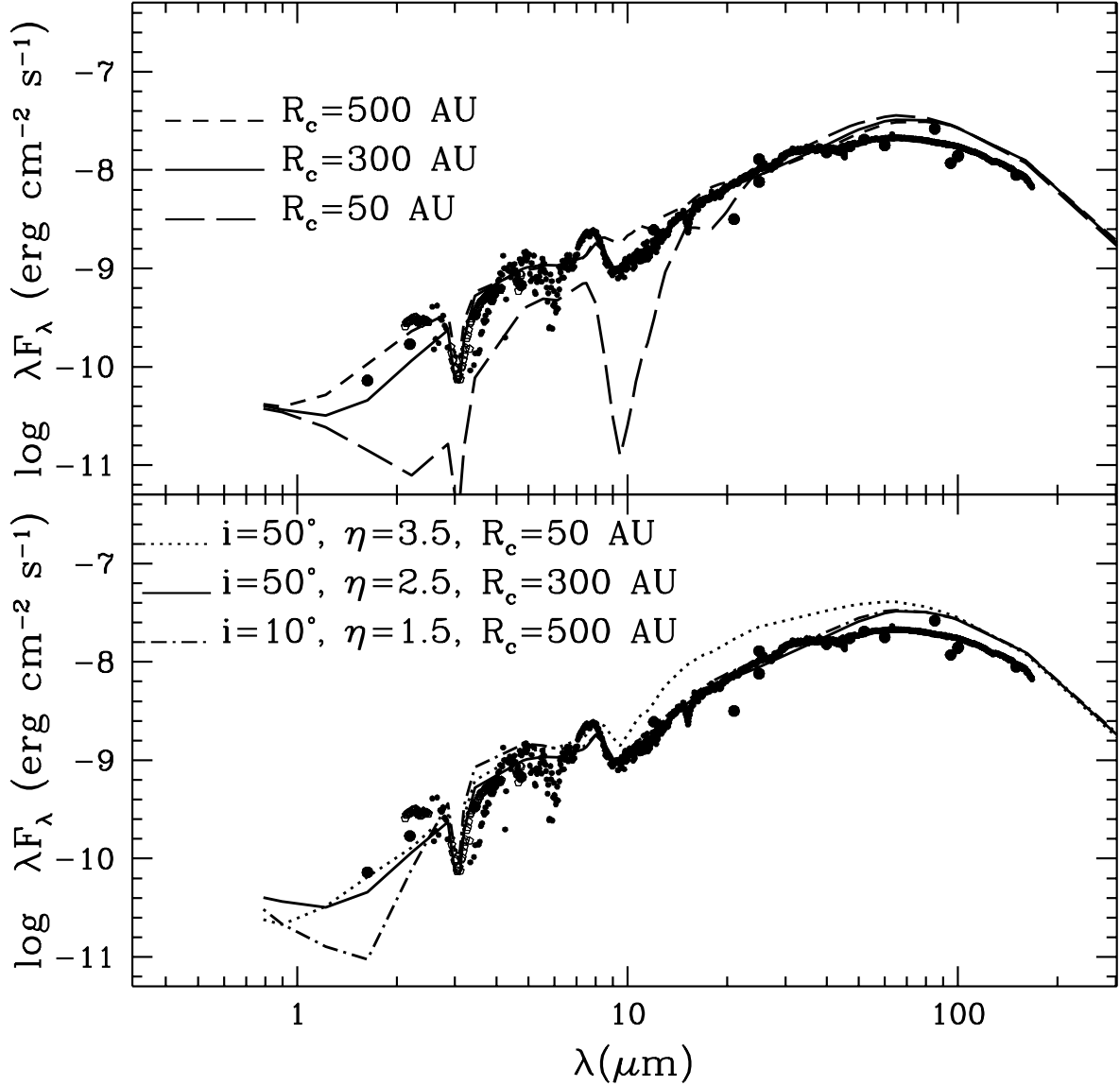}
\caption{Upper: SED  of the model  with $L=25 L_{\odot}$, 
 $\rho_1=4.5 \times 10^{-13}$ g cm$^{-3}$ and $\eta=2.5$ for 
  different centrifugal radii: 50 AU (long-dashed line), 300 AU (solid, the
 fiducial model) and  500 AU (short-dashed line). Observations as in Figure 
 \ref{grid1}. Lower: Comparison of model SEDs for high $\eta$, low $R_c$
(dotted) and low $\eta$, high $R_c$ (dot-dashed line) at inclinations that
 provide the best fit to the observations. The fiducial model 
(solid) is shown
for reference.} 
\label{gridrc}
\end{figure}

The upper panel of Figure \ref{gridrc} compares observations 
 with model SEDs  calculated for $i \sim 50^{\circ}$,  $\eta \sim 2.5$
and $R_c$ = 50, 300, and 500 AU.
All these models, with the same $\rho_1$, result in similar 
millimeter and centimeter fluxes because at these wavelengths 
the envelope is optically thin and the emission comes
from large scales. However, the
near and mid-IR fluxes are much more sensitive
to $\eta$, $i$, and $R_c$. The model with small centrifugal radius has too
much extinction along of the light of sight resulting in too deep
silicate and ice features; the model with large $R_c$ produces too little
absorption.  The parameters $\eta$, $R_c$, and $i$
are somewhat interchangeable (HCB96);
for instance, both an increase in $\eta$ or in $R_c$
result in low mean densities in the inner envelope.
However, there are differences. As $R_c$ increases, material tends to
pile up at larger distances from the luminosity
source on the equatorial plane; conversely, if $R_c$
is small, even if the envelope is very flat there is material near
the source and thus heated to higher temperatures. 
 So, model envelopes with large $\eta$ and small $R_c$ tend to have higher
near and mid-IR fluxes than low $\eta$ models with large
$R_c$, as shown in the lower panel of
Figure \ref{gridrc}. The parameters are thus
fairly well constrained when aiming to fit a large
range of wavelengths.

\subsubsection{Effect of dust opacities on the envelope SED} 
\label{efectopa}

One of the main uncertainties in determining the emission of YSOs lies in
the dust opacity, since the dust properties are not well known.  Several
models for the dust opacities have been used so far in the interpretation
of the SED of L1551 IRS 5, from interstellar medium dust (KCH93) to
mixtures of segregates and aggregates grains, the latter with cores of
silicates covered by ice mantles (White et al. 2000).  They all are
adjusted to fit the observations, so it is still not clear which is the
best representation of the grains for this object.
 
In this work, we compare two mixtures that have
been widely quoted in the literature of YSO in recent years
and determine which gives the best fit to the
observations. We calculate models for the mixture proposed by
P94 for molecular envelopes, which includes olivine,
organics, troilite, and water ice, with mass 
fractional abundances with  respect to gas of
$\zeta_{sil}=0.0034$, $\zeta_{org}=0.0041$,
$\zeta_{tro}=0.000768$,  $\zeta_{ice}=0.0012$.
We also use the DL84 mixture,
which characterizes 
the interstellar medium; this mixture consists of
grains  of graphite and ``astronomical'' silicates with 
a mass fractional abundances
of $\zeta_{sil}=0.004$ and $\zeta_{gra}=0.0025$.
We have added water ice to the DL84 mixture, to
account for the observed water ice bands, with
abundances similar or lower than that in the P94
mixture, as required to fit the observations. However,
  our mixtures  do not consider the ice 
 feature at 6 $\mu$m for the lack of optical constants in this
 band.   Furthermore, in the case of
DL84 mixture we    increase slightly the silicate abundance 
 ($\sim$ 20 \%) to improve the fit at the near IR.

\begin{figure}
\plotone{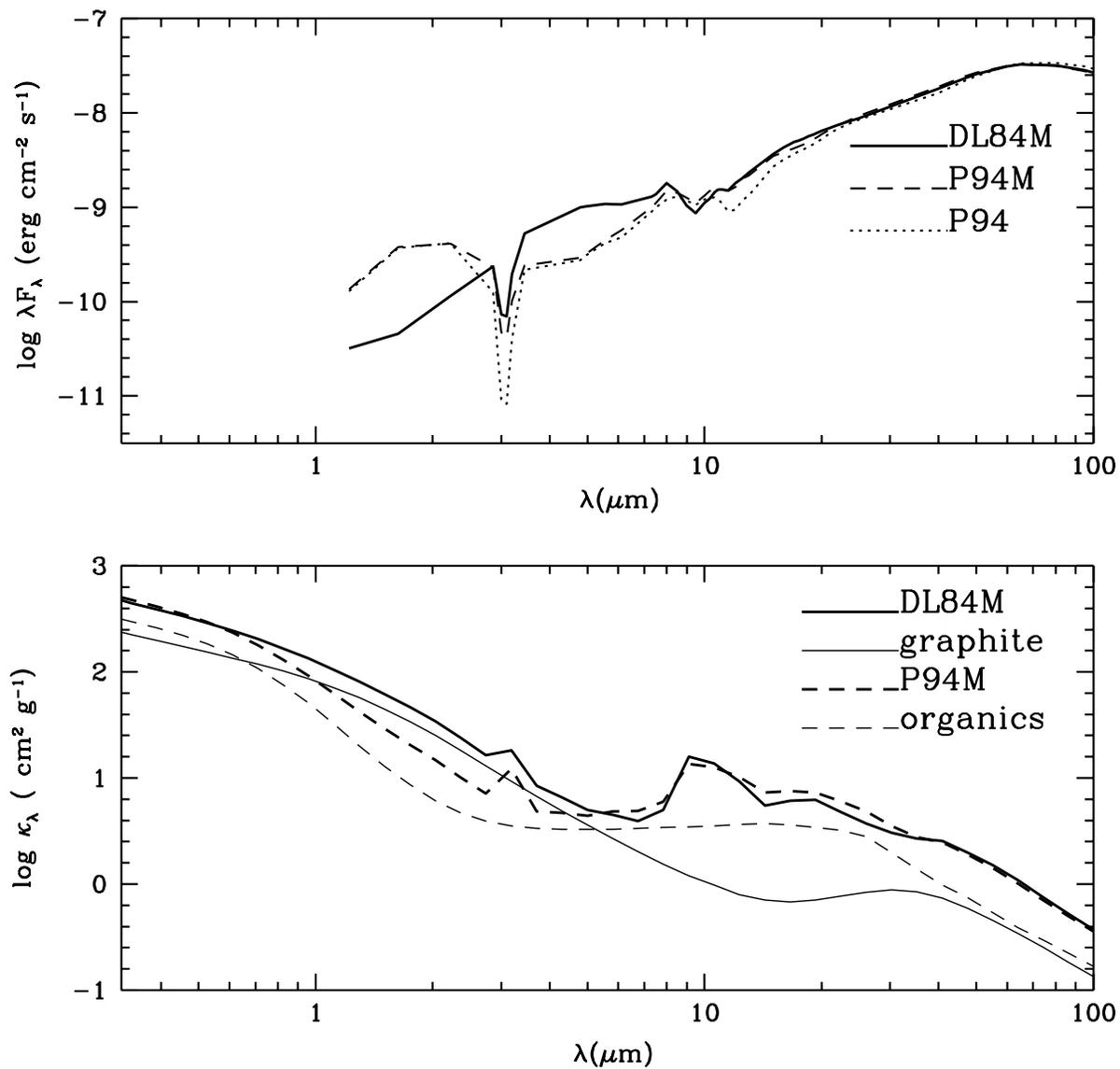}
\caption{ Upper: SED of the fiducial model with different dust mixtures.
 P94 (dotted line), P94M (dashed line), DL84M (solid line).
Lower: total mass absorption coefficient from P94M  (thick dashed line), 
 organics (light dashed line), total mass absorption coefficient 
 from DL84M (thick solid line), graphite (light solid line).} 
\label{grid1a}
\end{figure}

The upper panel of Figure \ref{grid1a} shows the  model SED
calculated for the parameters of the fiducial model
with three different dust mixtures: P94, P94 but with ice with  
approximately  half  of the abundance in P94, which we call P94M, 
and DL84, again with ice with  approximately half of the abundance 
in P94, which we call DL84M.
As discussed in \S \ref{envpar}, the model SED with 
DL84M at $i \sim 50^{\circ}$, gives the best fit to the
near to far infrared SED.
The water ice abundance is so high using P94
that the feature at 3 $\mu$m is  very strong, 
 much more than in the observations 
 and a feature of water ice at 12 $\mu$m appears which  
is also inconsistent  with the observations.
However, even with a reduced water ice abundance,
the P94M SED is different from the DL84M SED:
the near-IR are higher, while the mid IR flux  is 
lower which is in conflict with the observations.

The lower panel of Figure \ref{grid1a} compares
the P94M and the DL84M opacities.
 One important difference between the two mixtures
occurs in the near infrared range ($\sim 1 \mu$m - 5$\mu$m),
where the total opacity is mainly due to the
carbon component of the mixture and this element can be either in
 graphite or organics. Although the abundance of organics (P94M) is higher
 than that of graphite (DLM84), its extinction in this interval 
 is lower  than  that of  graphite 
 (see the behavior of the curves 
 of these components in Fig. \ref{grid1a}). 
Since the opacity is lower in the P94M mixture, 
the inner hot regions are more exposed with this  
mixture and the emergent flux is higher around $2 \mu$m.
However, also because the opacity is lower in the near-IR
range, where most of the light from the
circumstellar disks heating the envelope arises,
the P94M inner envelope has a temperature  $\sim 0.8$ times cooler 
 than the DL84M envelope.
In the mid-IR, the opacities for both
mixtures have more similar values, so
the depths where $\tau_{\lambda} \sim 1$ are approximately
the same. Since the same spatial regions
are contributing to the emission, but they are cooler for P94M,
the emergent flux is lower.
Because of these differences, we could not find any combination
 of parameters ($i, R_c, \eta$) that fitted the SED using   
 the P94 mixture, even with lower ice content.

\subsection{Millimeter Spatial Intensity Profiles}
\label{milsp}

In this section, we consider the density distribution at large
scales, inferred from the millimeter spatial intensity profiles
in the SCUBA images of Chandler \& Richer (2000, CR00).
A comparison in two dimensions between the SCUBA images 
of CR00 and our predicted images 
is not possible because the JCMT beam is asymmetric and time variable along 
the observing run (see CR00 for details).
Therefore, we averaged both the model and the data in circular annuli
centered at the peak position. This yields a mean spatial intensity profile,
$I(p)$ where $p$ is the angular radius from source center. We simulate
the observation at each wavelength convolving the  output model  with the  
 approximate beam  given by CR00;
the model includes the envelope, the circumstellar disks, and the 
 circumbinary disk.  These components are calculated with the parameters 
that give the best fit to other observables, as determined in \S
\ref{envpar}, \ref{diskresult}, and \ref{circumresult}.

\begin{figure}
\plotone{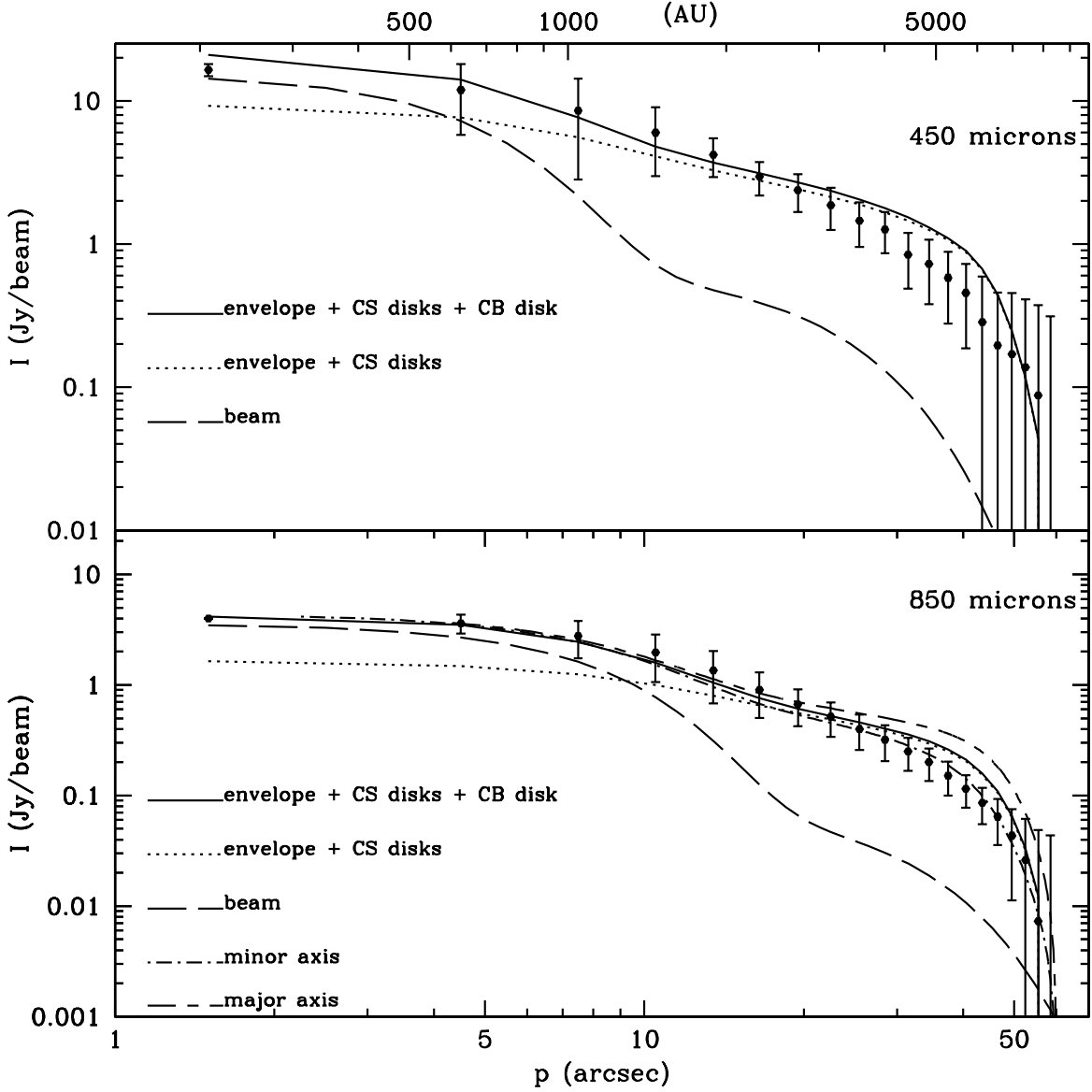}
\caption{
Spherically averaged spatial intensity profile at $450 \mu m$ (upper panel) 
 and $850 \mu m$ (lower panel). SCUBA
observations of  Chandler \& Richer 2000 (filled dots with error bars);
 predicted emission from the envelope (\S \ref{envpar}) 
plus circumstellar disks (CS, \S \ref{diskresult}) (dotted line);
predicted emission from the envelope, circumstellar disks,
and circumbinary disk  (CB, \S \ref{circumresult}) (solid line).
 The predicted profile at $850 \mu m$ along the
 minor axis (dot-dashed line) and the
 major axis (long-short dashed  line) are also shown.
The shape of the SCUBA beam is represented in each panel
(long-dashed line).  }
\label{profiles}
\end{figure}

Figure \ref{profiles}  compares the observed and the predicted 
spatial profiles at 450 and 850   $\mu$m, with observational errors 
 from CR00. Following CR00, we subtracted the intensity at 60" from 
both the observations and the model.
The dotted line shows the predicted emission without
the circumbinary disk; the solid line includes
all components of the system, showing the importance of the
circumbinary disk at small and intermediate scales.
The combined model provides a good fit to the observations,
specially for 850$\mu$m, which has the highest signal-to-noise
(CR00). 

To indicate the effects of flattening in the envelope,
we show in Figure \ref{profiles} model intensity profiles
calculated along the minor (rotational) axis and the
major (equatorial) axis of the envelope. The fact that
the observed profile is closer to that of the lower intensity 
minor axis suggests that there is less material
at large scales than expected, even for  a flat envelope,
 which may be due to the effect of the wind/jet.

\subsection{Circumstellar disk SEDs, 7 mm images and near IR CO absorption}
\label{diskresult}

The millimeter observations at high angular resolution can be used to fit
the properties of the circumstellar disks, since the contribution of the
envelope and circumbinary disk are resolved out (Rodr\'\i guez et al. 1998
and references therein;  Looney et al. 1997; also see Table 1), and
long-wavelength data indicate that contamination by free-free emission
from ionized outflows is unlikely (Rodr\'\i guez \etal~1998.)

Since we are assuming that the system luminosity arises in the
disks, we can equate $L_{tot}/2$ to the accretion luminosity
of each disk, $L_{acc}=G M_* \mdot /R_*$. Another
constraint comes from the maximum  effective temperature in the disk
for a steady viscous disk model,  
$T_{max}=0.29 \,({GM_*\mdot }/ {{{R_*}^3} \sigma})^{1/4}$,
which we can equate to $\sim 5000$~K, from estimates of the
optical scattered light spectrum (Stocke \etal~ 1988).
Combining these two constraints, we obtain
$R_*=1.4 R_{\odot}$ and 
$(M_*/0.3 \msun) (\mdot/ 2 \times 10^{-6} \msunyr) \sim 1$.
We adopt $M_* \sim 0.3 M_{\odot}$, 
close to the value expected for a typical low-mass T Tauri star
near the birthline (Stahler 1988; Hartmann, Cassen, \& Kenyon 1997),
so $\mdot \sim 2 \times 10^{-6} \msunyr$.
(Note that for an $\alpha$ disk, the surface density
$\Sigma \propto \mdot / \nu \propto \mdot {M_*}^{1/2}$, 
so we need to specify $\mdot$ and $M_*$ separately).

\begin{figure}
\epsscale{0.90}
\plotone{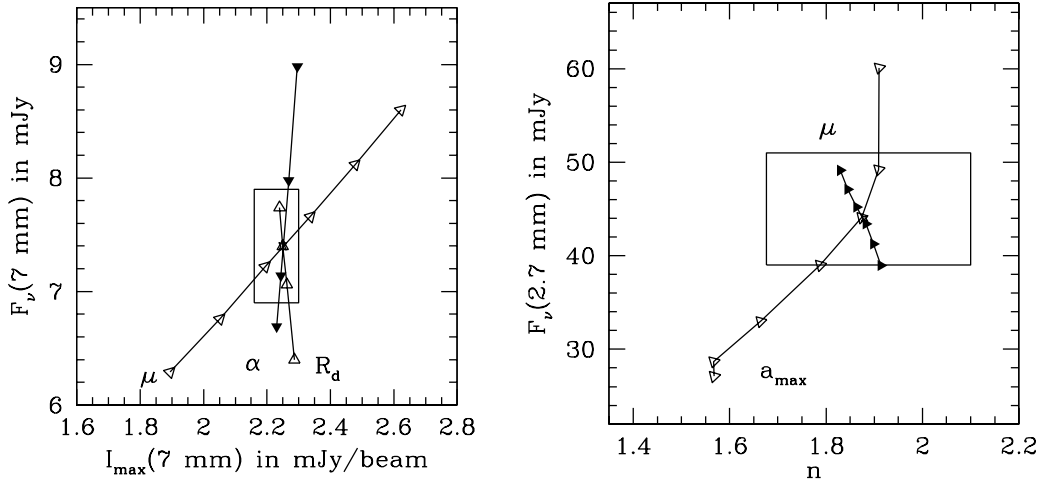}
\caption{
Dependence of observables on irradiated accretion disk parameters.
Left panel:  Flux vs maximum intensity at 7 mm as a function
of inclination angle $i$ (measured by $\mu =\cos i$), viscosity parameter
$\alpha$ (or $\sim$ disk mass) and disk radius $R_d$. Values
  plotted (increasing as
indicated by the arrows) are $\mu=0.4,0.45,0.5,0.55,0.6$ and 0.65; 
 $\alpha =$  0.0005, 0.001, 0.002,  and 0.003, 
corresponding to masses, $M_d=$0.7, 0.4, 0.23 and 0.16 $M_\odot$,
respectively; $R_d = $ 12, 13, 13.5 and 14 AU. Right panel:
Flux at 2.7 mm vs mm slope between 2.7 and 7 mm, as a
function of inclination and  $a_{max}$.
Values of $\mu$ are the same as in the left panel,  and
$a_{max}$ = 10, 100, 200, 300, 400, 500, and 600 $\mu$m.
The observation for the northern disk and its  errors 
(taken from Rodr\'\i guez et al. 1998) are  represented by a box.}
\label{diskvar}
\end{figure}

Figure \ref{diskvar} shows the dependence of observables on the disk model
parameters. On the left panel, we show the flux at 7 mm vs $I_{max}$,
which is the maximum intensity of the disk model image at this wavelength,
calculated by convolving the emergent intensity with a $0\rlap.''062
\times 0\rlap.''052$ gaussian beam. The observed values of these
quantities and the beam size are taken from Rodr\'\i guez et al. (1998).
Results are shown for different inclination angles (measured by $\mu =\cos
i$), viscosity parameter $\alpha$ (or $\sim$ disk mass), and disk radius
$R_d$, with values indicated in the figure caption. In the right panel of
Figure \ref{diskvar}, we show the total flux at 2.7 mm vs the slope
between 2.7 mm and 7 mm, defined as
$n=-[\log(F_{\rm 7mm})-\log(F_{\rm 2.7mm})]/[\log(7)-\log(2.7)]$ for 
different inclination angles and grain maximum sizes.

The mm flux increases with disk radius $R_d$, as expected, but the maximum
intensity is almost constant with $R_d$ because $I_{max}$ arises in the
inner $\sim $7 AU, for a $\sim  0\rlap.''05$ beam.  The flux also 
 increases when
the inclination angle decreases ($\mu$ increases), because these disks are
optically thick in the mm wavelength range; these are bright compact
disks, with smaller sizes and larger mass accretion rates than the typical
Classical T Tauri disks, usually optically thin at radio frequencies.

\begin{figure}
\epsscale{1.0}
\plotone{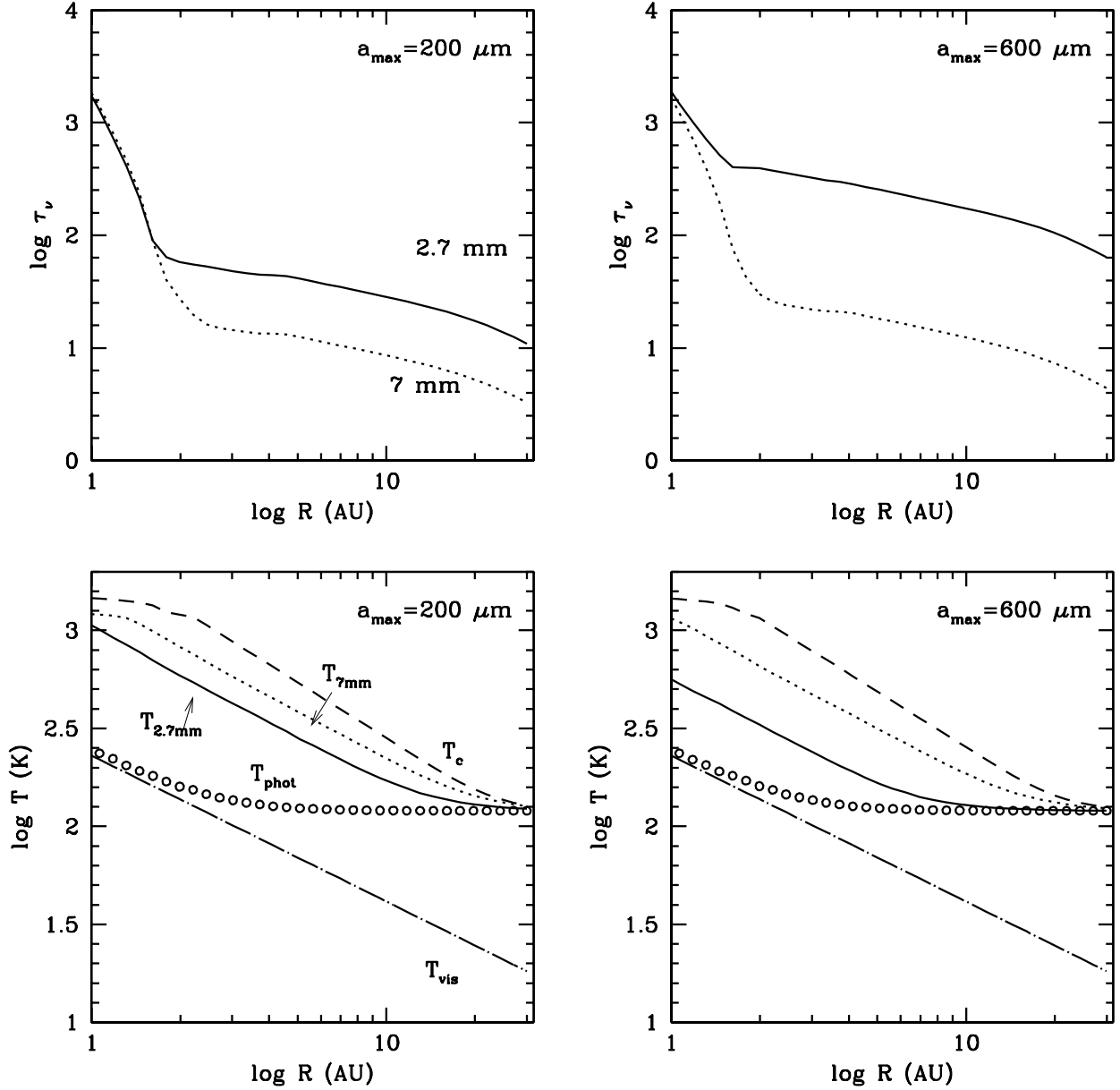}
\caption{
Properties of the circumstellar disks as a function of $a_{max}$.
Upper panels. Vertical optical depths as a function of radius
for 2.7 mm (solid) and 7 mm (dotted). The value of $a_{max}$ is indicated 
 in the panels. Lower panels.
Characteristic temperatures as a function of radius:
midplane temperature $T_c$ (dashed line), photospheric temperature $T_{phot}$
(circles). The temperatures of the height where the
optical depth at a given wavelength (for a pole-on disk) 
becomes unity: 2.7 mm (solid) and 7 mm (dotted) are also shown.
These correspond approximately to the brightness temperatures
at those wavelengths  (\S \ref{diskresult}).
Structural parameters other than $a_{max}$ as in Table \ref{tbl1}.
}
\label{diskdetails}
\end{figure}

The dependence of the emergent mm flux on the other
parameters of the model also reflects optical depth effects.
The upper panels of Figure \ref{diskdetails}
show the disk total vertical optical depth, $\tau_{\nu}$, at 2.7 mm and 7 
mm for a$_{max}$ 
equal to 200 $\mu$m (left column) and 600 $\mu$m (right column).
At both wavelengths, the optical depth is greater than 
one.
The lower panels of Figure \ref{diskdetails}
show relevant temperatures for the disk; in particular,
$T_c$ is the temperature at the midplane, $T_{phot}$
is the photospheric temperature, and 
$T_{vis}$ is the effective temperature corresponding to viscous
heating alone (DCH97). In a very rough approximation,
$T_{phot}^4  \sim T_{vis}^4 + T_{irr}^4$, where $T_{irr}$
is determined by envelope irradiation. At small radii,
$T_{phot} \sim T_{vis}$, but as radius increases, $T_{phot}$
approaches $T_{irr} \sim$ 120 K, the irradiation
temperature provided by the fiducial envelope model (\S \ref{envpar}).
 On the other hand, the Rosseland mean optical depth for $R \lesssim $ 13 AU 
in these disks is so high that the  midplane temperature is not 
very sensitive to envelope irradiation; it is mostly determined by local
viscous energy trapped close to the midplane (D'Alessio et al. 1999),
as $T_c \propto T_{vis} \tau_R^{1/4}$, where $\tau_R$ is
the vertical Rosseland mean optical depth, in the diffusion approximation.

The temperatures at which the optical depths at 2.7 mm and at 7 mm become
$\sim 1$ are also indicated in Figure \ref{diskdetails}.
These temperatures would be approximately the brightness temperatures
at those wavelengths if the disks were pole-on, 
so they reflect the characteristic emergent intensity.
 As already pointed out by Rodr\'\i guez et al. (1998), 
at mm-wavelengths the disk radiation emerges from a depth 
 close to the midplane. As a consequence of this, the disk emission
is characterized by a temperature  close to $T_c$
 which is much higher than the disk photospheric temperature 
because these are very optically thick disks.
This explains why the disks have a brightness temperature higher
than expected for the photosphere of a viscous disk, even with a high
mass accretion rate  $\sim  2 \times 10^{-6} \ \msunyr$ and
  illustrates the importance of constructing
detailed physical vertical structure models reflecting
the actual temperature gradients, and also the power
of interferometric imaging in the mm range to probe
the disk interior.

Figure \ref{diskvar} shows that the mm fluxes decrease
with increasing $\alpha$. The reason for this is that the
surface density of an $\alpha$ viscous disk
can be written as $\Sigma \propto \mdot / \alpha T_c$,
so in the diffusion approximation, $\Sigma \propto {\alpha}^{-4/5}$
and $T_c \propto {\alpha}^{-1/5}$. As $\alpha$ decreases,
the midplane temperature and thus the flux increase, although 
 slightly; the flux only increases by a factor of 
$\sim$ 1.3 as $\alpha$ decreases by a factor of $\sim$ 6. 
 Given the low dependence of 
$T_c$ on $\alpha$, the mass of the disk
 varies  approximately as $\alpha^{-1}$; we have indicated in the
figure the disk mass that corresponds to each of the
models shown.

The mm fluxes and slope depend on the dust opacity,
and in particular, on the maximum grain size $a_{max}$, although
the disks are optically thick. This is a consequence of the
fact that as  the dust opacity changes, it affects both  
  the height in the disk where $\tau_{\nu} \sim 1$,
and  the vertical thermal distribution (in the regions of the disk
where the opacity is dominated by dust.)
The larger the $a_{max}$, the smaller the $\tau_R$,
 which is similar to the optical depth in the mid-IR
for the range of temperatures in the disk (cf. DCH01); 
thus, as $a_{max}$ increases,  the midplane temperature decreases.
In contrast, the mm opacity increases with $a_{max}$ (cf. DCH01).
 However, the increase is larger for wavelengths around 2.7 mm
for the range of sizes shown in Figure \ref{diskvar}.
This is because spherical grains, with a real refraction index $n > 1$,
show an opacity enhancement around $\lambda \sim 2 \pi <a>$, for
an average grain radius $<a>$
(cf.  Miyake \& Nakagawa 1993; P94; DCH01). 
This effect produces a larger contrast in opacity between 2.7
and 7 mm when $a_{max}$ 
increases, in the range of sizes considered, which
then implies a larger contrast between the
temperatures $T(\tau_\nu \sim 1)$  at both wavelengths
as $a_{max}$ increases. The difference in $T(\tau_\nu \sim 1)$ 
 explains why the slope of the mm SED decreases as $a_{max}$ increases.
The consideration of other kind of grain shapes or porosity
will tend to reduce or eliminate this enhancement of opacity
(e.g., Miyake \& Nakagawa 1993)
 making  the effect of $a_{max}$ on the slope of the mm SED less noticeable.

Figure \ref{diskvar} shows boxes 
centered in the observations of the
northern disk with sides given by the observational uncertainties.
 A similar analysis can be made for the southern disk.  The
resulting model parameters adopted for each disk are shown in 
Table \ref{tbl1},
with uncertainties given by the ranges analyzed in Figure \ref{diskvar}.

\begin{figure}
\plotone{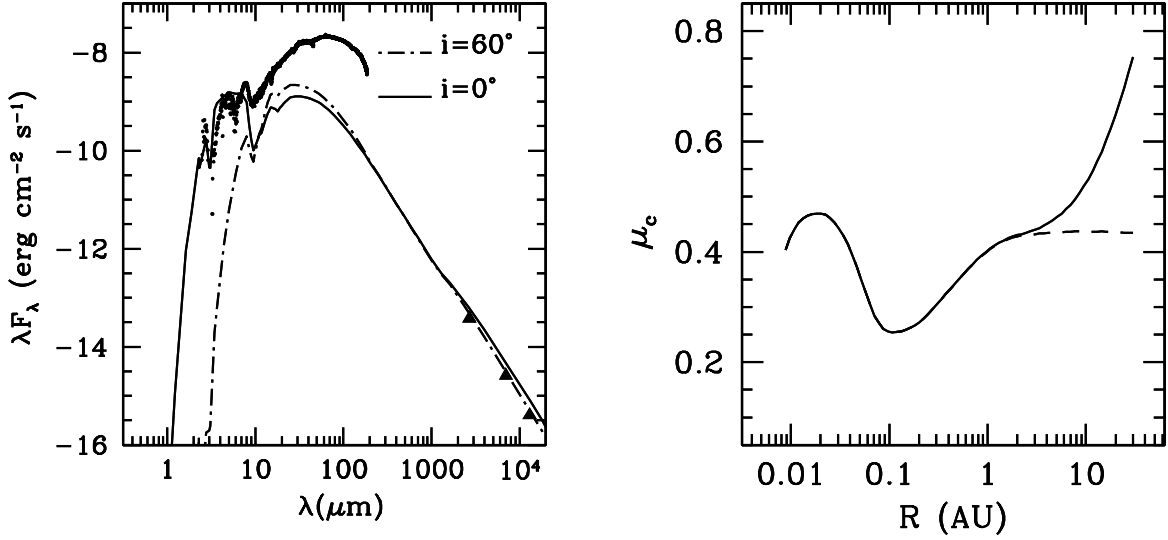}
\caption{ 
 Left panel: SED for an irradiated disk model with parameters in Table
\ref{tbl1}, with an inclination angle $i=0^{\circ}$ (solid line)  and $i =
60^{\circ}$ (dotted-dashed line), showing the effect of self-occultation.  
Interferometric measurements of the northern disk (triangles;
Rodr\'{\i}guez et al. 1998 and references therein) and the ISO data (dots)
are also shown. Right panel: Cosine of the critical angle, $\mu_c$, as a
function of disk radius for radiation characteristic of the inner disk
($T\sim 4500$ K). For $\mu < \mu_c$, the outer disk absorbs the radiation
of the inner disk. The dashed line is the purely viscous disk, the solid
line is the disk irradiated by the envelope (\S \ref{diskresult}).
 } 
\label{disks}
\end{figure}

The SED of an individual circumstellar disk for inclinations
$i = 0^\circ$ and $i = 60^\circ$ is shown in the left panel of
Figure \ref{disks}, as well as the interferometric
data for the northern disk from Looney et al. (1997)  and 
Rodr\'\i guez et al. (1998). The ISO data is shown as a reference.
The SED has been attenuated by the optical
depth of the envelope model that produces the best fit to the
   IR SED (\S 5.1). At low inclinations,  disk
emission results in too much near-IR. However,  at 
inclinations that produce the best fit to the mm data 
 (cf. Figure   \ref{diskvar}), the near-IR emission of the disks 
drops because the disks are so flared 
 that the outer regions of the disks attenuate the
inner disk emission along the line of sight.   
To illustrate this, we show in the right panel of Figure \ref{disks}
the cosine of the critical inclination angle $\mu_c(R)$, 
 such that the inner disk is attenuated for values of
inclinations $\mu < \mu_c (R)$, for a disk of radius $R$. 
 The critical angle is defined as $i_c = \cos^{-1} \mu_c $,
such that $\mu_c = z_c/(z_c^2+R^2)^{1/2}$, and 
the  height $z_c$ is that where the optical depth along a
ray from  the star to a point in the disk at radius $R$ is unity.
The optical depth is evaluated at a wavelength where radiation
from the inner disk, with $T \sim 4500 \ K$, is expected to peak.
 Two disk models are shown in the right panel of Figure \ref{disks}, 
 a purely viscous  disk and a viscous disk irradiated by 
 the envelope, with parameters
that provide the best fit to the mm data. 
The self-occulting effect is 
much more important in the case of the irradiated               
disk because  the atmosphere at large radii is  much hotter and
has a larger scale height than the pure-viscous case;
specifically, the atmosphere of the  outer disk occults the inner
disk when its axis is inclined by  $ \mu <  0.4$ ($i > 66^\circ$)
to the line of sight for the non-irradiated disk 
and by $\mu < 0.6$ ($i > 50^\circ$) for the irradiated disk 
 (for $R_d \sim 10$ AU).
For the case $i = 60^\circ$ shown in Figure \ref{disks},
the optical depth due to the disk along a ray that passes through
the center of the star is $\tau(1 \ \mu m) \sim 25$.
This self-occultation of the disk, arising as a consequence
of other observational constraints, explains nicely the
fact that at the near infrared, the observed emission
is extended (Lucas \& Roche 1996). In our interpretation, the near-IR
emission is mostly light from 
the inner disk scattered out by the outer disk and the
surrounding envelope towards the observer, and thus it is seen as 
an extended rather than a pointlike source.

\begin{figure}
\plotone{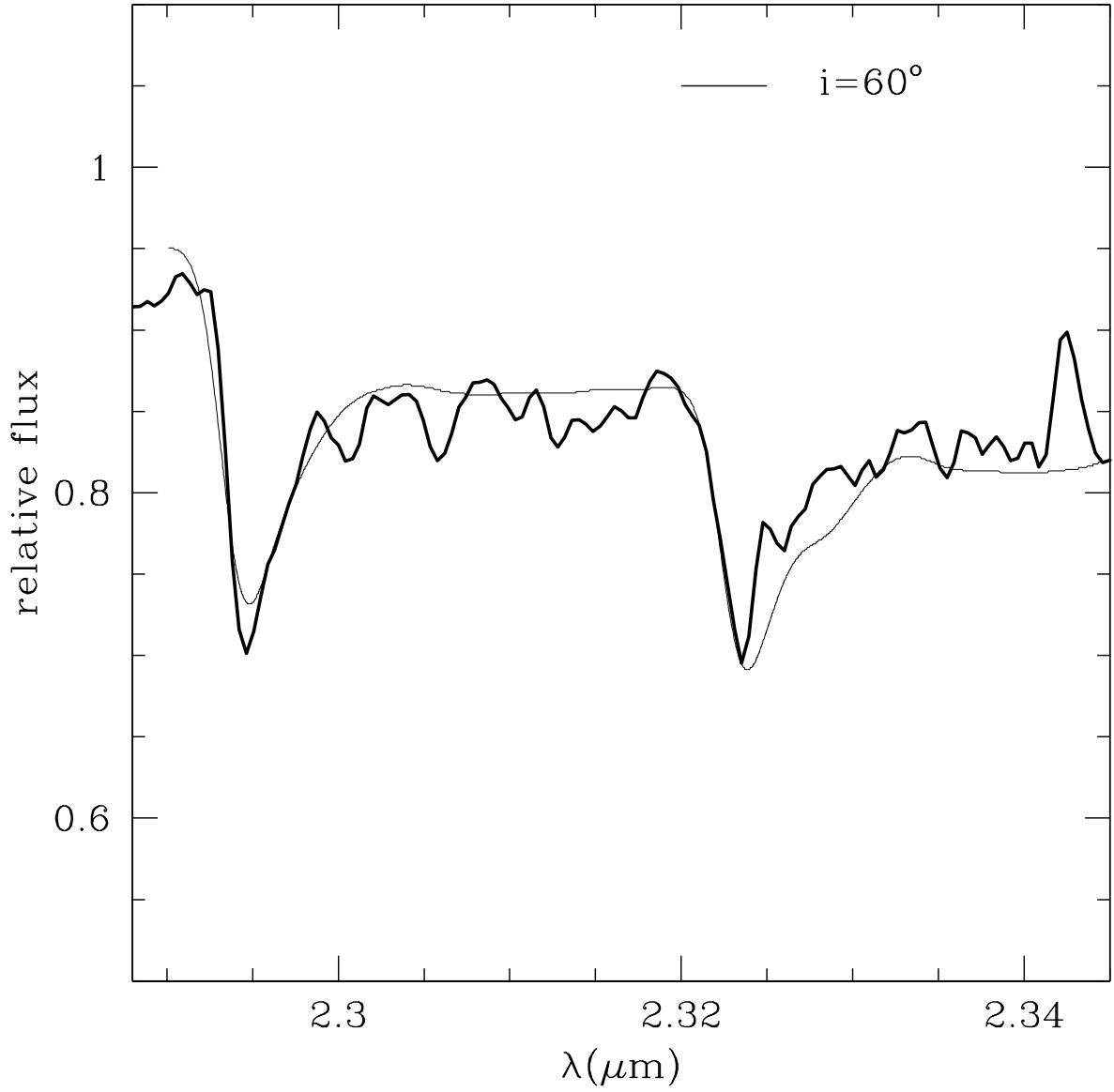}
\caption{Portion of our SpeX spectrum of L1551 IRS 5 showing the
first overtone bands of CO (heavy line). Calculations
for a disk of mass accretion rate $2 \times 10^{-6}$ M$_{\odot}~yr^{-1}$,
stellar mass $0.3 M_{\odot} $ and stellar radius $1.4 R_{\odot}$,
at an inclination angle of 60$^{\circ}$ (light line). The calculations
have been convolved to a resolution of R=2000, and
both observations and predictions have been normalized
to the continuum.
}
\label{co}
\end{figure}

We have calculated the flux of the first overtone
CO bands in the near infrared, formed in a disk atmosphere  
whose density and temperature profiles are determined using the same 
mass accretion rate 
and stellar parameters as each of the binary
disks, following the procedures in Calvet et al. (1991). 
Figure \ref{co} shows the result of these 
calculations for an inclination angle $\sim$ {60$^\circ$},
convolved with a resolution of R=2000. At this resolution, 
disk spectra for inclinations less that $\sim$ 70$^\circ$
would be similar. Disk models
with parameters that reproduce the long wavelength, 
 the near-infrared and total emission constraints
 predict very well the observed strengths of the
first two CO first overtone bands in our SpeX spectrum,
as shown in Figure \ref{co}, 
providing an additional confirmation of the disk structural
parameters.

Since the inner disks are occulted, the flux at
2.3$\mu$m is actually radiation from the inner disks
scattered by dust at the edges of the disks and in the envelope (a 
similar conclusion was reached by White et al. 2000).
A contribution from thermal emission of the hot dust in the
innermost regions of the envelope should be added to the light scattered
from the disk, which would tend to decrease the
strength of the absorption; as more material
is located in these hot regions, the stronger the
veiling expected in the CO bands, as discussed in Calvet, Hartmann,
\& Strom (1997). As shown in that paper, the veiling
roughly scales with the density at the dust
destruction radius. With a mean density of 
  $\sim 4 \times 10^{7} {\rm cm^{-3}}$ 
from our envelope calculations, we obtain a veiling
$\sim 0.3$ scaling from their results. This is likely to be an upper
limit, since we expect more evacuation in the $\eta$
models than in the TSC models used in Calvet, Hartmann,
\& Strom (1997), which is consistent with the 
low veiling of the CO bands indicated in Figure \ref{co}.

\subsection{Circumbinary Disk SED}
\label{circumresult} 

As discussed in \S \ref{sistema}  and  \S \ref{modeldiscogrande},  
the observations of Keene \& Masson (1990),  Lay et al. (1994), 
 Hogerheijde et al. (1997)
 suggest dust emission with a scale of 100-400 AU.
Looney, Mundy \& Welch (1997) also found evidence of a 
 structure of similar size from  emission at 2.7 mm and they estimated that 
this structure could be a circumbinary disk  
with a mass of $\sim 0.04 M_{\odot}$. 
Following these suggestions, we have included a circumbinary disk in
our model. The outer radius of this disk is limited
by the centrifugal radius, as discussed in \S \ref{modeldiscogrande};
from the result of our envelope SED modeling, \S \ref{envpar}, we infer
$R_c \sim $300 AU, approximately in  agreement with expected sizes.
We have taken the inner disk radius as 120 AU, three times the apparent 
separation between the circumstellar disks.

\begin{figure}
\plotone{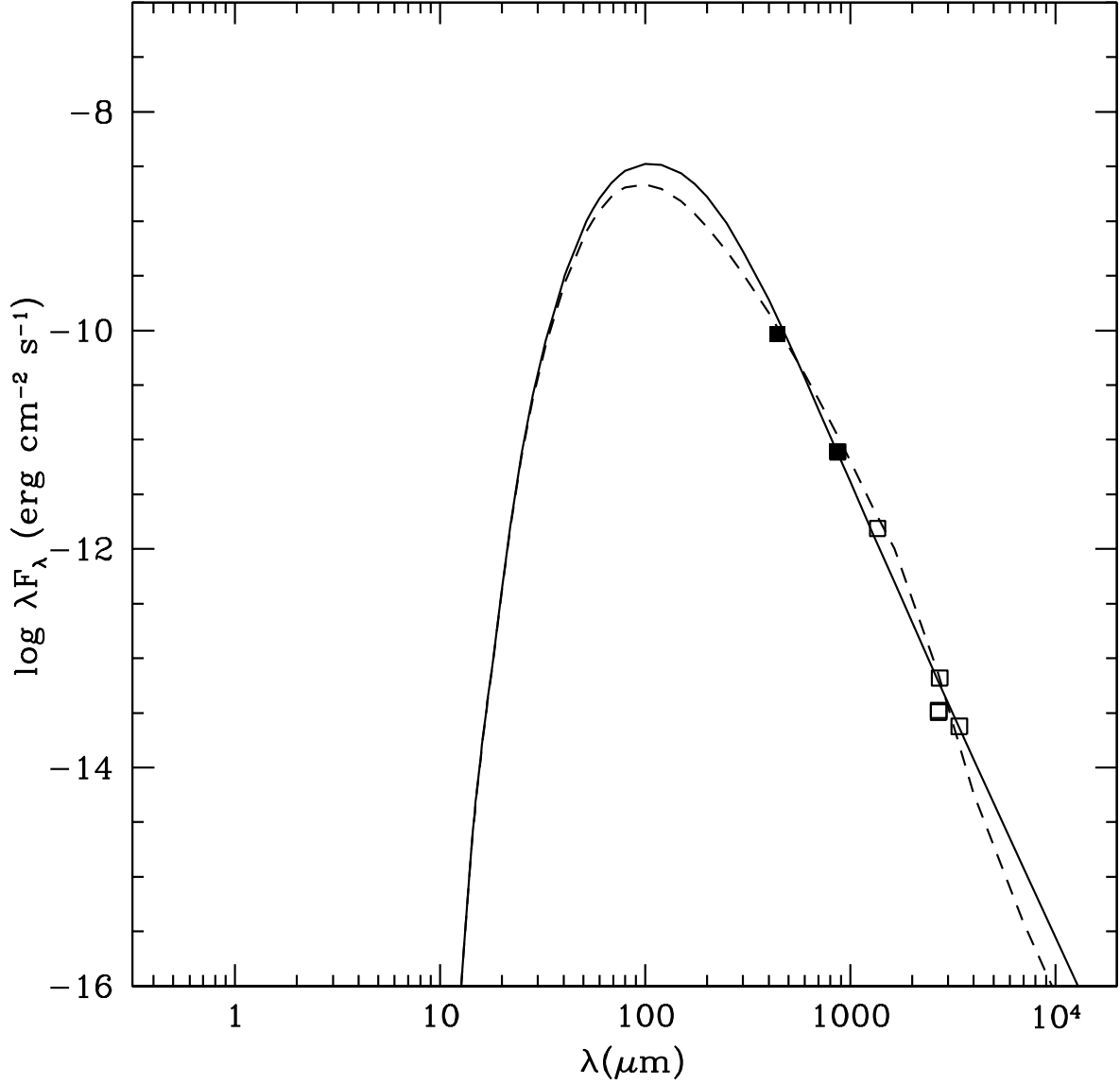}
\caption{
 SEDs of the circumbinary disk obtained assuming a dust size distribution
as in the envelope (solid line) or as in the circumstellar disks (dashed
line).  Fluxes at 450~$\mu$m and 850~$\mu$m (filled squares) come from the
fit to the large-scale intensity distribution profiles, \S \ref{milsp}.
Fluxes at other wavelengths are listed in Table 1 (open squares).
 }
\label{CB}
\end{figure}

Figure \ref{CB} shows the fluxes for this structure.
Fluxes at 450 $\mu$m and 850 $\mu$m are  the fluxes added to the envelope to 
 fit large-scale intensity distribution profiles, \S \ref{milsp}.
 The latter flux agrees with the flux obtained by Lay et al. (1994).
Fluxes at 2.7 mm have been estimated by Keene \& Masson (1990),
 Looney et al. (1997) and  Hogerheijde et al. (1997) and the flux at 
 3.4 mm has been estimated by Hogerheijde et al. (1997). 
 For these fluxes   we have subtracted the contribution
from the circumstellar disks (\S \ref{diskresult}) since at 2.7 and 3.4 mm
  their emission is substantial.

 The irradiation flux produced by the  fiducial model of \S \ref{envpar}, 
 yields a temperature at 100 AU of $T_0 \sim$ 55K.
 With this temperature, we model the fluxes in Figure \ref{CB} to obtain
the density distribution scaling $\Sigma_0$; this value is
very sensitive to the adopted opacity. 
Figure \ref{CB} shows fits for a dust size distribution 
similar to that in the envelope ($a_{max} = 0.3 \mu$m) and
a dust size distribution as in the circumstellar disks 
($a_{max} = 400 \mu$m). The mass surface density at 100 AU for these
models is $\Sigma_0 = 34~ {\rm g~ cm^{-2}}$
and $\Sigma_0 = 1.3~ {\rm g~ cm^{-2}}$, respectively,
with corresponding masses of 0.4 and 0.02 $\msun$. 
 The  large difference in disk mass is due to the drop of mm opacity
 as a$_{max}$ decreases. The mass obtained by Looney et al. (1997) is 
 near our lower limit, because they use an opacity consistent with 
large grains.  Our models for the circumbinary disk are little 
restricted, however we favor the model obtained with the envelope dust,
 since material from the envelope is arriving on the equatorial plane of 
the system mostly near $R_c$, raining on top of the circumbinary disk.
 Interferometric observations of high rotational molecular line
transitions (such as those that will be achievable with the SMA in the
near future), that would trace the warm component of the gas, together
with a more detailed modelling of these line emission could strengthen the
case for the existence of the circumbinary ring and determine better its
physical parameters.

\subsection{Summary of results}

\begin{figure}
\epsscale{0.90}
\plotone{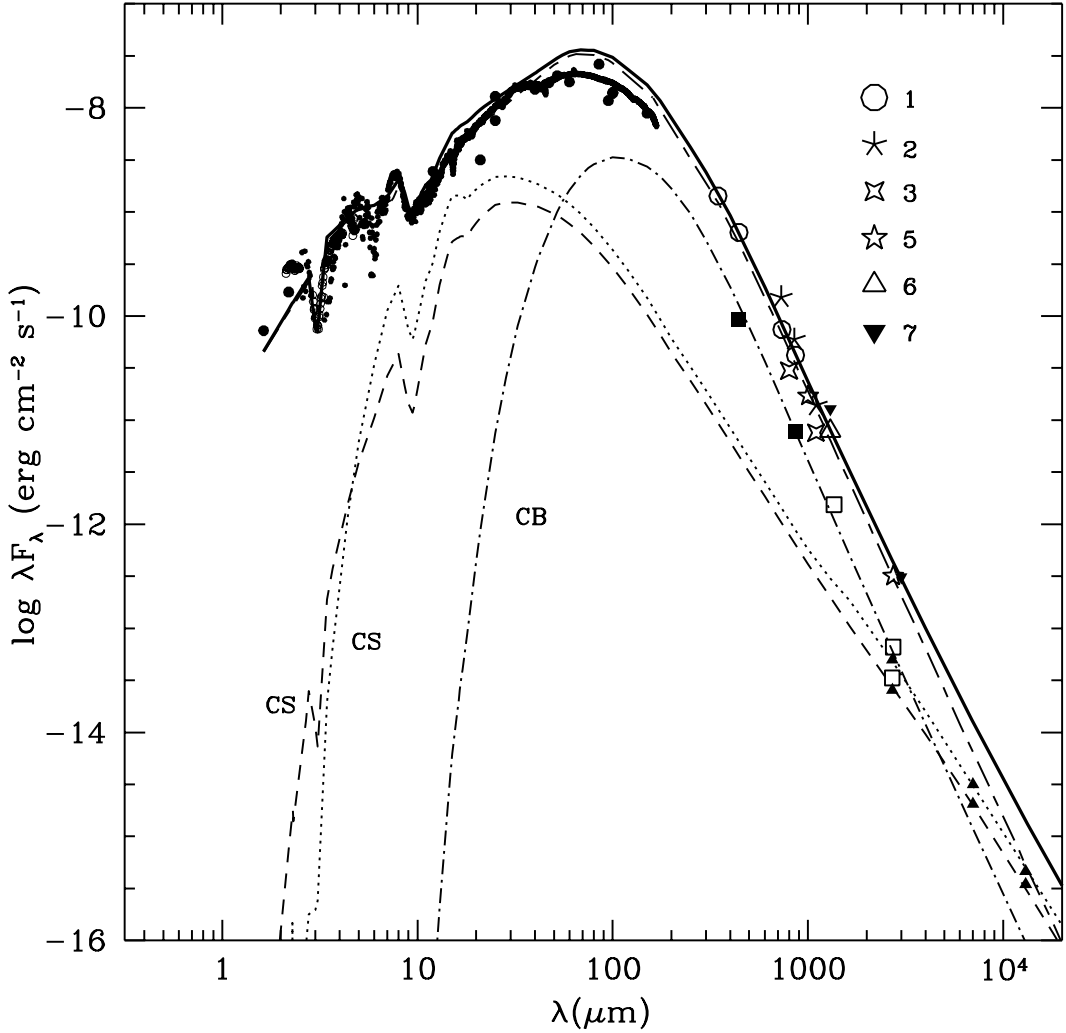}
 \caption{ Composite SED (envelope + circumbinary disk + circumstellar
disks) for a model with the parameters listed in Table \ref{tbl1} (solid
line). Predicted SED of the envelope (short and long-dashed line).
Predicted SEDs of the circumstellar disks (CS): Northern component (dotted
line) and Southern component (dashed line).  Predicted SED for a
circumbinary disk (CB, dotted and dashed line). The IR observational data
represented are from ISO (filled-small dots), SpeX (open-small dots), and
photometry observations compiled by KCH93 (filled-big dots). Data for the
circumstellar disks (filled triangles) and for the circumbinary disk
(filled and open squares) and are the same as in Figures \ref{disks} and
\ref{CB}, respectively. For the remaining data points, the number from 
Table 1 reference list is indicated.
 } 
\label{seds}
\end{figure}

To summarize, the far-IR SED
fluxes the density in the envelope, while the degree of flattening, the
centrifugal radius and the inclination are mostly determined
by the near and mid-IR fluxes. The submillimeter images
  also constrain the density of the envelope,
  at scales of hundreds of AU, they require an additional contribution
from the circumbinary disk. The millimeter interferometric
images  and fluxes, as well as the total luminosity of the system,
the optical and near-IR spectra, and the near-IR images
determine the properties of the circumstellar disks.
The circumstellar disks are the source of the luminosity that
heats the envelope; in turn, envelope irradiation
is a key element in the heating of the disks in the
system. Figure \ref{seds} shows the overall fit to the data, indicating
each component separately. The parameters characterizing
each component are given in Table \ref{tbl1}.

\section{DISCUSSION}

One of the challenges to models posed by recent observations is the
detection of giant envelopes (as determined from observations by Chandler
\& Richer 2000 for L1551 IRS 5, and by Shirley et al. 2000, Motte \&
Andr\'e 2001, Larsson, Liseau \& Men'shchikov 2002 in other protostars)
combined with a limited depth in the silicate absorption feature (White et
al. 2000). As discussed in \S 5, previous envelope models predict silicate
features that are too deep (Butner \etal~1991;  KCH93) even with small
envelopes; increasing the envelope limits makes the problem worse.  The
flattened density distribution of the sheet collapse or $\eta$ model
alleviates this problem by reducing the amount of extinction along the
line of sight to the central region.  As observations suggest that the
envelope of IRS 5 is indeed flattened, this solution to the silicate
feature problem seems reasonable.

The value of ice abundance (H$_2$O/H  $\lesssim~4 \times 10^{-5}$) require
 to fit the envelope SED, so that the absorption to 
 3 microns is reproduced and that at the same time  the absorption
 of water ice at 12 microns does not appear, is a factor of two lower 
 than  determinations  of P94 and  
 Tielens et al. (1991). This latter author determined approximately 
 the abundance in L1551 IRS 5 from the optical depth
 of water ice, silicates features and assuming standard dust-to-gas 
 correlations. With  our abundance  we estimated a water ice column density of
 $\sim~1.5 \times 10^{19}$ cm$^{-2}$ at 
 the radius where the   sublimation temperature occurs
  ( $\sim 100~K$, for water ice cf. Sandford \& 
Allamandola 1993 ) in our best fit to the envelope SED.  However,
a determination of the column density deduced  from the optical  
depth at  3.08 microns from  our spectrum using $I=I_oe^{-\tau}$
and $N_{dust}$(H$_2$O)$ =\tau_{3.08} \Delta \nu_{1/2}/A $
(Lacy et al. 1984; Sandford et al. 1988), where $A$ is the
band strength and $\Delta \nu$  is the full-width at half maximum intensity,
 provides a  lower value ( $\sim~3.8 \times 10^{18}$  cm$^{-2}$, consistent 
 with  that found by Tegler et al. (1993), and  within 
a factor of two of that determined by  White et al. 2000, 
 using the same method). The discrepancy with our value inferred
 from the best fit might be caused by  determinations  of  the optical depth
  from $I=I_oe^{-\tau}$, where $I_o$ and $I$ are the intensities
measured from the spectrum at continuum level and at 
the position of the ice absorption respectively,   can underestimate
 the optical depth and then the column density 
by ignoring local emission in the radiative transfer.
  Thus, one must to take the value of the column density inferred 
 in this way only as a lower limit. 
 
We also found that the P94 mixture with no graphite but organic grains was
not able to fit simultaneously the near-infrared and millimeter range of
the spectrum. While we find reasonable fits to the SED with the modified
Draine \& Lee opacities, we do not therefore conclude that graphite must
necessarily be present, merely that the overall opacity curve must
resemble the results for this mixture.  Other models, for example those of
Li \& Greenberg (1977) with organic refractory mantles for silicate core
grains, might also work well.

We require slightly different inclination angles for the accreting
disks at the center of the IRS 5 system in comparison with the
inclination adopted for the envelope.  However, the difference
is small, considering the assumptions and simplifications involved
in our calculations.  It is also quite possible that the jets and
therefore their originating disks are not perfectly aligned  
(Rodr\'\i guez \etal~ 1998).
The circumstellar disks are quite massive (see Table \ref{tbl1}) compared to 
typical Classical T Tauri disks. The disk models are  gravitationally 
unstable against axisymmetric perturbations for $R >$ 6 AU 
 and $R > 8$ AU, for the northern and southern disks respectively,
and  the Toomre parameter 
at $R_d$ is  $Q_T (10 AU) \sim 0.5$ and $0.8$, for each disk. 
Also  the interaction between both disks and them with the circumbinary 
ring could strongly affect the disk structure
and the pertinence of the $\alpha$ prescription can be questioned. 

In our model, the envelope material falls mostly near the centrifugal
radius, which is considerably outside the positions of the two disks.
Using the value of $\rho_1$ for the envelope,
the envelope infall rate is 
 $\sim 7 \times 10^{-5}\, (M_{sys}/0.9 \msun)^{0.5}\, M_{\odot} yr^{-1}$, 
 where $M_{sys}$ is the mass at the center, i.e., the sum
 of stellar and disk masses. Notice that the value of $M_{sys}$ in 
 our model is roughly  consistent with 
the dynamical mass of the binary system recently estimated by 
 Rodr\'\i guez et al. (2002). Thus, the  mass infall rate of the envelope is 
  an order of magnitude or more  larger than the total accretion rate through
the two disks, $\sim 4 \times 10^{-6} M_{\odot} yr^{-1}$.  While these
disk accretion rate estimates are somewhat model dependent, they are consistent
with the total system luminosity; if the disk accretion rates were really
as large as $7 \times 10^{-5} M_{\odot} yr^{-1}$, the total system luminosity
for the protostellar mass-radius relation used above (roughly birthline
values; Stahler 1988) would be roughly 
 $L_{acc} = G \mdot M_*/R_* \sim 440 \lsun$,
an order of magnitude larger than observed.  This suggests
that material may be piling up, for example in the circumbinary ring
or disk suggested by the interferometric observations at intermediate
resolutions (\S 2,  \S \ref{circumresult}), and is not currently 
 accreting in the disks.
KCH93 noted a similar ``luminosity'' problem for Taurus Class I sources;
in general, the infall rates estimated from the TSC models would produce
accretion luminosities an order of magnitude higher than observed if
the infalling material accreted at the same rate onto the central protostar.
KCH93 also suggested that material is piling up onto outer disks without
immediately and continuously accreting inward.  In the case of IRS 5,
we have for the first time additional confirmation of this suggestion, 
in that the disk accretion rate directly estimated from the mm observations 
is much lower than the infall rate.

It is unreasonable to suppose that material can pile up indefinitely in
an outer circumbinary disk, at least when that disk accumulates a mass
comparable to that in the central regions.
Following the proposal of Kenyon \& Hartmann (1991), KCH93 suggested that
FU Orionis outbursts occur when disk material that has accumulated in
outer radii becomes gravitationally unstable and catastrophically
accretes in a burst.  The FU Ori characteristics
seen in the optical and near-infrared spectra features (Carr \etal 1987;
Stocke \etal 1988) provide additional circumstantial evidence for this
model.  Circumbinary disk mass estimates are strongly dependent
upon the assumed opacities, for a $M \sim 0.02-0.4 \msun$ given 
from our fits,
 the disk  may be gravitationally unstable  supporting the fact
 that a FU Ori event is happening.
The FU Ori outburst may be  ending, with the circumbinary disk already
having been emptied out;
the current IRS 5 accretion rates are several times smaller than the
typical FU Ori accretion rates at maximum, $\sim 10^{-4} M_{\odot} yr^{-1}$
(Hartmann \& Kenyon 1996), consistent with a considerable decline from
maximum accretion rate.  In any event, our results emphasize the
need to consider protostellar evolution with variable accretion rates,
and the probable importance of differing disk accretion rates and
envelope infall rates for studying evolution and fragmentation in
protostellar disks.

\section{ SUMMARY AND CONCLUSIONS}

Our main results can be summarizes as follows:

\begin{enumerate}

\item The wealth of observational data available for L1551 IRS 5, 
 in particular the observed SED  by ISO and the water ice feature  at 
 3 microns by SpeX has enabled us to develop 
 detailed models for the 
 circumstellar envelope,  constraining its luminosity, geometry,  
infall accretion  rate, and  opacity.

\item We find that flattened collapse models with an inclination along 
 the line of sight $\sim 50^{\circ}$, a total luminosity of 
 $\sim 25~L_{\odot}$, mass $\sim 4~ M_{\odot}$, 
  infall accretion rate  $\sim$ 7$\times 10^{-5}$  
 $M_{\odot}~ yr^{-1}$ and a normal opacity laws, modified
by changing the water ice abundance, enable us to fit simultaneously 
 the ISO, SpeX  observations and  single dish submillimeter and
  millimeter fluxes  
 that are mostly dominated  by the emission of the large scale envelope.

\item We  find that the a mixture as the proposed by Pollack et al. 1994
 with no graphite but organic grains was not able to fit
simultaneously the near-infrared and millimeter range of the spectrum.
While we find reasonable fits to the envelope 
 SED with the modified Draine \& Lee
opacities, we do not  conclude that graphite must necessarily
be present, merely that the overall opacity curve must resemble the
results for this mixture.

\item  At smaller  scales than 2000 AU, the millimeter spatial 
 intensity profiles in the SCUBA images suggest  the presence of the 
 circumbinary disk. We model this structure with a grain size distribution 
similar to that in the extended envelope and that in the circumstellar 
disks, as the two extreme possibilities for the state of the dust in 
this disk. We find that the mass of the 
 circumbinary disk goes from 0.02 to 0.4 $M_{\odot}$ for $a_{max}=400$
 $\mu$m and $a_{max}=0.3$ $\mu$m, respectively.

\item The properties of the circumstellar disks are inferred from high 
angular millimeter images and fluxes, optical spectra and near-IR image. 
 We find that these disks 
are optically thick in the 
mm wavelength range; and  are inclined  an angle ($\sim 60^{\circ}$),
  so that  their outer parts hide  the emission from their inner parts.
 We also find that these disks have larger mass accretion rates 
 ($\sim  2 \times 10^{-6} \ \msunyr$) than the typical 
Classical T Tauri disks, usually optically thin at radio frequencies.

\item The millimeter emission of the circumstellar disks emerges from a depth
 close to the midplane. As a consequence of this, the disk emission
is characterized by a temperature   
 much higher than the disk photospheric temperature. 
This explains why the disks have a brightness temperature higher
than expected for the photosphere of a viscous disk, even with a high
mass accretion rate   and illustrates the importance of constructing
detailed physical vertical structure models reflecting
the actual temperature gradients, and also the power
of interferometric imaging in the mm range to probe
the disk interior.

\item The circumstellar disk models
with parameters that reproduce the observational constraints, 
 predict very well the observed strengths of the
first two CO first overtone bands detected in our SpeX spectrum,
 providing an additional confirmation of the disk structural
parameters.

\item In the case of L1551 IRS 5,
we have for the first time the confirmation of the suggestion, 
that material is piling up onto outer disks without
immediately and continuously accreting inward since
 the disk accretion rate directly estimated from the mm observations 
is much lower than the infall rate,
probably resulting in the accumulation of material in a circumbinary disk,
possibly with occasional cascades of accretion into the binary system 
to produce FU Orionis eruptions.

\end{enumerate}

\acknowledgments

{\it Acknowledgments}

 We thank Claire Chandler for providing us the original SCUBA data of
L1551 IRS 5 and Mario van den Ancker for help with the ISO data.
 Thanks as well to Robert Estalella and Maite Beltr\'an for providing a
program to convolve our results with the observing beam of SCUBA camera.
 We also thank Guillem Anglada for useful discussions and helpful
suggestions.
 J.M. would like to thank Bobby Bus and John Rayner for support at IRTF.
 M. O.  acknowledges the fellowship from Conacyt, M\'exico and
 the NASA Origins of Solar Systems grant NAG5-9670.  
P. D. acknowledges the grant J27748E from Conacyt, M\'exico.
We thank our referee for his/her thoughtful review of our manuscript and 
useful comments.

\clearpage

\begin{deluxetable}{cccccc}
\tabletypesize{\scriptsize}
\tablecaption{\sc Compilation of submm and mm Observational Data  
\label{datos}}
\tablewidth{0pt}
\tablehead{
\colhead{$\lambda$}
&\colhead{}
&\colhead{Angular}
&\colhead{Flux}
&\colhead{} 
&\colhead{}\\
&\colhead{}
&\colhead{Resolution}
&\colhead{Density}
&\colhead{Size}
&\colhead{}\\
\colhead{($\mu$m)}
&\colhead{Instrument }
&\colhead{($''$)}
&\colhead{(Jy)}
&\colhead{(AU)}
&\colhead{Refs.}
}
\startdata
350   &JCMT     & 12  &164  & $\sim$ 8000      &        1\\
450   &JCMT     & 11  & 94  & $\sim$ 8000      &        1\\
730   &JCMT      & 16  & 37     & $\sim$4200   &        2\\
730   &JCMT      & 16  & 3           & $\sim$1300   &   2\\
750   &JCMT     & 14    & 18  & $\sim$ 8000    &        1\\
800   & JCMT   & 17    & 8             & \nodata     &  3\\
850   &JCMT     & 15  & 12  & $\sim$ 8000      &        1\\
850   &JCMT      & 16  & 17          & $\sim$4200&      2\\
870   &JCMT-CSO  & $\sim$1-4 &  2.24   & $\sim$80   &   4\\
1000  & Hale Telescope    & 55     & 5.7    & \nodata   & 5\\
1100  &JCMT      & 19  &  5         & $\sim$5600&       2\\
1100  & JCMT   & 18    & 2.8           & \nodata    &   3\\
1300  &IRAM      & 11  &  3   &  $\sim$10000         &  6\\
1300  &12 m NRAO    &  $\sim$30  & 0.7  & \nodata     & 7\\
2700  &OVRO      & 3  & 0.097         & $<$~400 &       8\\
2700  &BIMA\tablenotemark{a}  & 0.3  & 0.045&    $<$~25  & 9 \\
2700  &BIMA\tablenotemark{b}  & 0.3  & 0.023&    $<$~25  & 9 \\
2730  &OVRO      & 3 &  0.29        & $\sim$2000      & 5\\
2730  &OVRO     &  3  & 0.13 &  $\sim$ 50      &        5\\
3000  &12 m NRAO    &  $\sim$60  & 0.7  & \nodata   &   7\\
3400  &OVRO      & 6  & 0.081         & $<$~400 &       8\\
7000  &VLA\tablenotemark{a} & 0.05 & 0.0074  & $\sim$10  & 10\\  
7000  &VLA\tablenotemark{b} & 0.05 & 0.0048  & $\sim$10 & 10\\ 
13000 &VLA\tablenotemark{a} & $\sim$ 0.1 & 0.0020 & \nodata &10\\
13000 &VLA\tablenotemark{b} & $\sim$ 0.1 & 0.0015 & \nodata &10 
\enddata
\tablenotetext{a}{Northern disk.}
\tablenotetext{b}{Southern disk.}
\tablerefs{
 (1) Chandler \& Richer 2000 (using SCUBA);
 (2)  Ladd et al. 1995 (using the UKT14 bolometer);
 (3) Moriarty-Schieven et al. 1994 (using the UKT14 bolometer);
 (4) Lay et al. 1994;
 (5) Keene \& Masson 1990;
 (6) Motte \& Andr\'e 2001; 
 (7) Walker. et al. 1990;
 (8) Hogerheijde et al. 1997; 
 (9) Looney et al. 1997;
 (10) Rodr\'\i guez et al. 1998 and references therein.
}
\end{deluxetable}

\begin{deluxetable}{lcc}
\tabletypesize{\scriptsize}
\tablewidth{0pt}
\tablecaption{\sc Parameters of the Model\label{tbl1}}
\tablehead{ & Envelope &
}
\startdata
L$_{tot}$ ($L_{\odot}$)  & 25 & \\
$i$ (degrees)  & 50 \\
$\rho_1$  (${\rm g~ cm^{-3}}$) & $4 \times 10^{-13}$&\\
{$\dot M_{infall}$} ($M_{\odot}~ yr^{-1}$) & 7 $\times 10^{-5}$&\\
$R_c$ (AU) & 300 & \\
$R_{out}$ (AU)   & $\sim 8000$ &\\
$\eta$  &2.5 & \\
$a_{max}$ ($\mu$m)  & 0.3 &\\
Mass ($M_{\odot}$)& 4  & \\
\hline
& Circumbinary Disk & \\
\hline
$i$ (degrees) & 50& \\
Inner Radius (AU) & 120 & \\
Outer Radius (AU)  & 300 & \\ 
$a_{max}$ ($\mu$m)  & 0.3 &\\
$\Sigma_0$ ($\rm g~ cm^{-2}$)& 34 &\\
Mass ($M_{\odot}$)& 0.4 & \\
$Q_T$(300 AU) & 0.01 &\\
\hline
& Northern Disk & Southern Disk \\
\hline
$M_*$ $(M_{\odot})$ &0.3           & 0.3 \\
$R_*$ $(R_{\odot})$ &1.4           & 1.4 \\
$i$ (degrees) & 60 & 62   \\
{$\dot M_{acc}$} ($M_{\odot}~yr^{-1}$)& 2 $\times 10^{-6}$& 
2$\times 10^{-6}$\\
$R_d$ (AU) & 13 & 12    \\
$a_{max}$ ($\mu$m)  & 200 & 400  \\
$\alpha$ & 0.0016 & 0.0047\\
$M_{d}$ ($M_{\odot}$)& 0.25 & 0.1\\
$Q_T$(10 AU) & 0.5 & 0.8 \\ 
\enddata
\tablecomments{Assumed distance is 140 pc.}
\end{deluxetable}

\clearpage

\end{document}